\documentclass[aps,prb,twocolumn,superscriptaddress,nofootinbib,longbibliography]{revtex4-2}
\usepackage[pdftex]{graphicx}
\usepackage{amsmath,amssymb,mathtools}
\usepackage[colorlinks=true,citecolor=blue]{hyperref}
\usepackage{amsfonts}
\usepackage{dsfont}

\usepackage{physics}

\usepackage[utf8]{inputenc} 
\usepackage[T1]{fontenc} 

\usepackage{xspace}

\usepackage[english]{babel}
\usepackage{lmodern}

\begin{document}

\title {Synchronized coherent charge oscillations in coupled double quantum dots}

\author{Eric~Kleinherbers}
\email{eric.kleinherbers@uni-due.de}
\affiliation{Faculty of Physics and CENIDE, University of Duisburg-Essen, 47057 Duisburg, Germany}
\author{Philipp~Stegmann}
\email{psteg@mit.edu}
\affiliation{Department of Chemistry, Massachusetts Institute of Technology, Cambridge, Massachusetts 02139, USA}
\author{J\"urgen~K\"onig}
\affiliation{Faculty of Physics and CENIDE, University of Duisburg-Essen, 47057 Duisburg, Germany}

\date{\today}

\begin{abstract}
We study coherent charge oscillations in double quantum dots tunnel-coupled to metallic leads.
If two such systems are coupled by Coulomb interaction, there are in total six (instead of only two) oscillation modes of the entangled system with interaction-dependent oscillation frequencies.
By tuning the bias voltage, one can engineer decoherence such that only one of the six modes, in which the charge oscillations in both double quantum dots become synchronized in antiphase, is singled out.
We suggest to use waiting-time distributions and the $g^{(2)}$-correlation function to detect the common frequency and the phase locking.
\end{abstract}

\maketitle

\section{Introduction}
Quantum coherence is the central ingredient in the development of new quantum technologies for computation, sensing, communication, imaging, and metrology~\cite{zagoskin_2011}.
However, coherent dynamics in a quantum device is typically tainted by the coupling to its environment. 
In this paper, we study a system constructed from two double quantum dots~\cite{takatura_2014,mukho_2018,banszerus_2020} and succeed in exploiting both tunnel-coupling to metallic leads as well as the Coulomb interaction to facilitate, rather than destroy, a clear and well-pronounced signal of synchronized coherent charge oscillations. 

\begin{figure}[t]
  \includegraphics[width=0.99\linewidth]{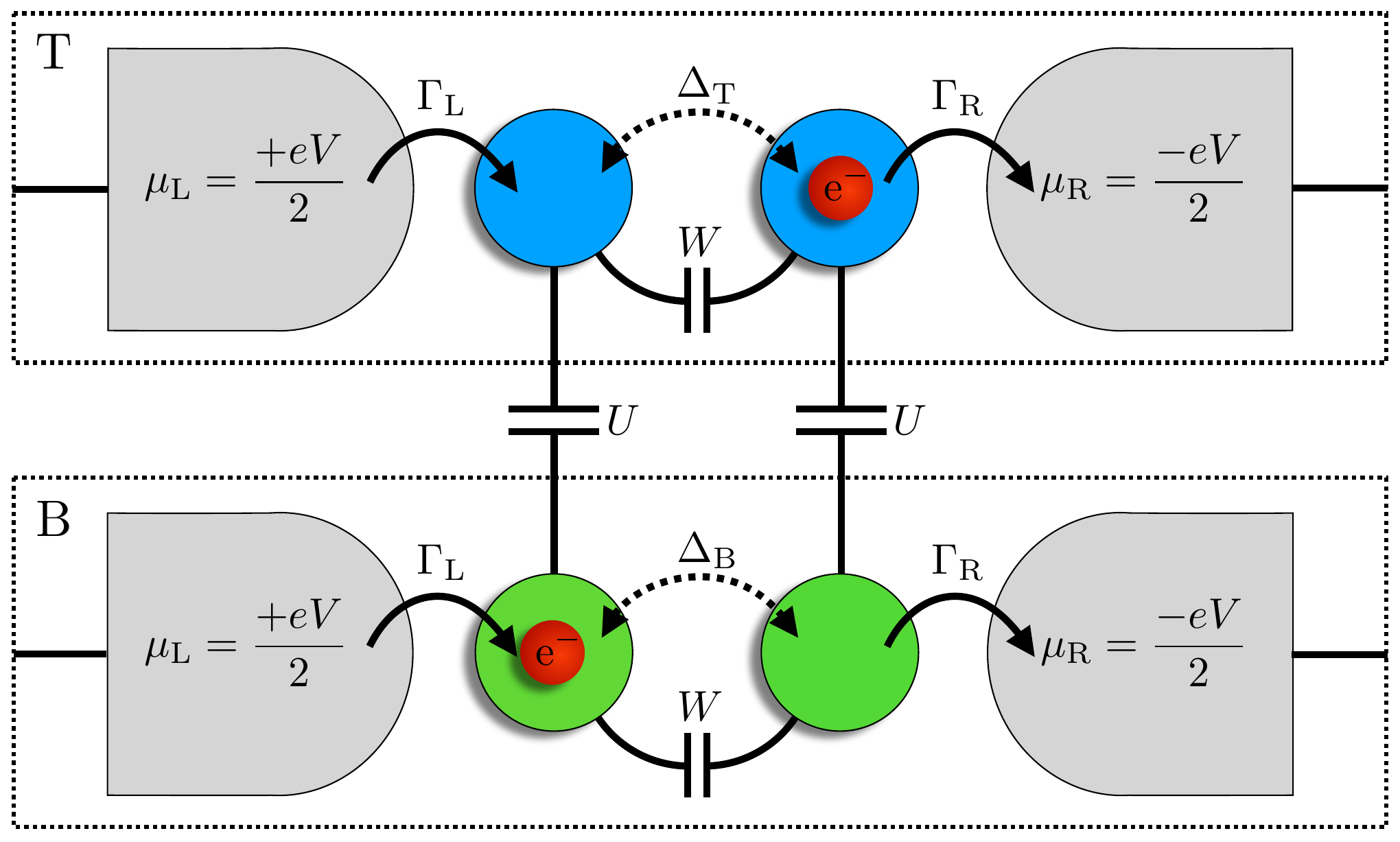}
  \caption{Model. The top DQD (blue) and the bottom DQD (green) are both tunnel coupled to a left and right metallic lead with tunnel-coupling strengths $\Gamma_\text{L}$ and $\Gamma_\text{R}\ll\Gamma_\text{L}$, respectively. The applied bias is $\mu_\text{L}-\mu_\text{R}=eV$. Once an electron tunnels into a DQD, it oscillates coherently between the left and right quantum dot before it tunnels out again. The parameters of the top and bottom DQD are identical except for the tunneling amplitude $\Delta_\text{T}<\Delta_\text{B}$. In addition, the DQDs are coupled via the Coulomb interaction $U$. }
  \label{fig:1}
\end{figure}

A generic example for coherent charge oscillations is the double quantum dot (DQD), where electrons coherently oscillate back and forth between the left and right quantum dot at a specific frequency~\cite{brandes_2008, kambly_2013, korotkov_2001,korotkov_1999, gurvitz_1997}. If two such DQDs with initially different oscillation frequencies are capacitively coupled~\cite{kushihara_2012,fujisawa_2011}, see Fig.~\ref{fig:1}, the systems become entangled~\cite{mansour_2020,filgueiras_2020,fanchini_2010} and instead of two different oscillation frequencies, a total of six are found. 
Here, we study the electron transport through such coupled DQDs when they are connected to left and right metallic leads with an applied bias voltage. 
We find an interesting regime where only one out of all six frequencies is singled out and the coherent charge oscillations in each of the two interacting DQDs become synchronized in antiphase.  This effect is caused by the interplay of entanglement induced by the Coulomb interaction and decoherence induced by charge fluctuations into and out of the metallic leads. In particular, a key mechanism for the synchronization is a non-equilibrium effect where the charging energy supplied by the coupling of the two DQDs is utilized to tunnel against the natural direction of the applied bias, so that coherent charge oscillations involving high-energy states can effectively decohere. 

Coherent charge oscillations in both a single DQD~\cite{kim_2015,srinivasa_2013,cao_2013,petersson_2010} as well as in two capacitively coupled DQDs~\cite{li_2015,shinkai_2009} have already been successfully measured by pump-probe type experiments, where the system is first manually excited via optical or electrical pulses and then read out. Here, however, we propose to perform a real-time measurement of the total number of electrons in each DQD (e.g., by a capacitively coupled quantum point contact~\cite{vandersypen_2004,gustavsson_2009} or single-electron transistor~\cite{fujisawa_2004,lu_2003})
so that not only the average current but all single-electron tunneling events are resolved. 
Coherences between the respective left and right quantum dots are preserved since only the total electron occupation of a DQD needs to be resolved in the measurement process.  
The benefit of such a real-time detection of single-electron transport is that it can be performed in a steady-state situation.
Nonetheless, by employing appropriate statistical tools as the \textit{waiting-time distribution}~\cite{brandes_2008,rajabi_2013,haack_2014,sothmann_2014,kosov_2017,kosov_2017_2,potanina_2017,rudge_2018,walldorf_2018,tang_2018,engelhardt_2019,rudge_2019,stegmann_2021,davis_2021,landi_2021} and the $g^{(2)}$-\textit{correlation function}~\cite{emary_2012}, one can effectively mimic a pump-probe experiment just by statistical means of analyzing the data. In particular, by evaluating waiting times between a tunneling-in event (pump) and a tunneling-out event (probe) coherent charge oscillations can be observed.  This offers the great advantage that no special initial state needs to be manually prepared and still information about the quantum dynamics can be extracted in real time.

This paper is organized as follows. In Sec.~\ref{sec:model}, we introduce the master equation for the studied system of two coupled DQDs connected to metallic leads. Then, in Sec.~\ref{sec:synchronization}, we show signatures in the waiting-time distribution and the $g^{(2)}$-correlation function indicating synchronized  coherent charge oscillations with a common frequency. In Sec.~\ref{sec:frequencies}, we investigate the isolated system of two coupled DQDs and find that there are in principle a total of six frequencies emerging from the Coulomb interaction between the DQDs. 
In Sec.~\ref{sec:decoherence}, 
we show that one of the six frequencies can be singled out if the decoherence mechanism induced by the metallic leads is suitably tuned by the bias voltage.
In this case, a synchronized coherent charge oscillation in antiphase can be observed. 
In Sec.~\ref{sec:conclusion}, we conclude our findings.

\section{Model}\label{sec:model}
The minimal model to study coherent charge oscillations in electron transport is a serial DQD tunnel-coupled to a left and right electronic lead~\cite{brandes_2008, ptaszynski_2017,thomas_2014,kambly_2013, korotkov_2001,korotkov_1999, gurvitz_1997}. In this paper, we examine two such systems -- a top and a bottom DQD (see Fig.~\ref{fig:1}) --- coupled by the Coulomb repulsion $U$ and study how the coherent charge oscillations affect each other. The top $(\alpha=\text{T})$ and bottom $(\alpha=\text{B})$ DQD are described by the Hamiltonians
\begin{align}
H_{\alpha}&=\frac{\varepsilon}{2} \left( d^{\dagger}_{\alpha,\text{L}} d^{\phantom{\dagger}}_{\alpha,\text{L}}- d^{\dagger} _{\alpha,\text{R}}d^{\phantom{\dagger}}_{\alpha,\text{R}}  \right)  \\ &-~\frac{\Delta_\alpha}{2} \left( d^{\dagger} _{\alpha,\text{L}}d^{\phantom{\dagger}}_{\alpha,\text{R}} + d^{\dagger}_{\alpha,\text{R}} d^{\phantom{\dagger}}_{\alpha,\text{L}}  \right) \nonumber \\ &+ ~ W ~n_{\alpha,\text{L}}n_{\alpha,\text{R}},\nonumber
\end{align}
where the fermionic operators $d_{\alpha,\beta}^{\dagger}$ and $d_{\alpha,\beta}^{\phantom{\dagger}}$ create and annihilate an electron in the quantum dot specified by $(\alpha,\beta)$, respectively. Here, $\alpha\in \{\text{T},\text{B}\}$ labels the top and bottom DQD and $\beta\in \{\text{L},\text{R}\}$ discriminates between the left and the right quantum dot. 
The occupation number operator is defined as $n_{\alpha,\beta}=d_{\alpha,\beta}^{\dagger} d_{\alpha,\beta}^{\phantom{\dagger}} $. The first two terms of the Hamiltonian describe the detuning $\varepsilon$ and tunneling~$\Delta_\alpha$ between the left and right quantum dot, respectively, whereas the last term describes the Coulomb repulsion $W$ within a DQD. 
Then, the full Hamiltonian can be written as
\begin{align}
\label{eq:Hdot}
H=H_{\text{T}}+H_{\text{B}}+U n_{\text{T},\text{L}}n_{\text{B},\text{L}}+U n_{\text{T},\text{R}}n_{\text{B},\text{R}},
\end{align}
where the charging energy $U$ only has to be paid if either both left or both right quantum dots are occupied with an electron. For simplicity, crossed capacitive couplings are neglected.

The DQDs are very weakly coupled to the electronic leads via tunnel barriers such that individual electrons tunnel sequentially into and out of the DQDs.
Then, the dynamics is governed by the Lindblad equation 
\begin{align}
\dot{\rho}&={\cal L}\rho=\frac{1}{i\hbar} \left[ H,\rho \right]  \nonumber \\&+ \sum_{\alpha,\beta,s} \Gamma_\beta \left( L^{\phantom{\dagger}}_{\alpha,\beta,s} \rho L^\dagger_{\alpha,\beta,s} - \frac{1}{2} \left\{L^\dagger_{\alpha,\beta,s} L^{\phantom{\dagger}}_{\alpha,\beta,s},\rho \right\} \right)
\label{eq:lindblad}
\end{align}
for the density matrix $\rho(t)$, where the indices run over $\alpha\in\{\text{T,B}\}$, $\beta\in\{\text{L,R}\}$ and $s\in\{+,-\}$. We set $\hbar=1$. 
While tunneling between the quantum dots is treated exactly with the Hamiltonian $H$, the coupling to the leads is treated perturbatively in the tunnel-coupling strengths $\Gamma_\text{L}$ and $\Gamma_\text{R}$. The coupling strengths are assumed to be equal for both DQDs and we define $\Gamma{=}\Gamma_\text{L}{+}\Gamma_\text{R}$. 
Finding the Lindblad operators that adequately describe the tunneling events into and out of the electronic leads is a nontrivial task. In a microscopic derivation, additional approximations are usually required to obtain a Lindblad form~\cite{kleinherbers_2020}.
When employing the widely-used secular approximation, each tunneling event into ($s=+$) or out of ($s=-$) the quantum dot $(\alpha,\beta)$ is described by multiple Lindblad operators $L_{\alpha,\beta,s}(\Delta E)$, one for each single-electron excitation energy $\Delta E$. 
As an improvement, we use here the so-called coherent approximation~\cite{gediminas_2018,ptaszynski_2019,kleinherbers_2020} instead, which leads to only one Lindblad operator describing the coherent sum $L_{\alpha,\beta,s}=\sum_{\Delta E} L_{\alpha,\beta,s}(\Delta E)$ over all excitation energies.
In particular, the Lindblad operators can be found via $L_{\alpha,\beta,+}=\sum_{\chi,\chi^\prime}\sqrt{f(E_\chi{-}E_{\chi^\prime}{-}\mu_{\beta})}\mel{\chi}{d_{\alpha,\beta}^\dagger}{\chi^\prime} \dyad{\chi}{\chi^\prime}$ and  $L_{\alpha,\beta,-}=\sum_{\chi,\chi^\prime}\sqrt{1{-}f(E_{\chi^\prime}{-}E_\chi{-}\mu_{\beta})}\mel{\chi}{d_{\alpha,\beta}}{\chi^\prime} \dyad{\chi}{\chi^\prime}$, where $E_\chi$ and $\ket{\chi}$ denote the eigenvalues and eigenstates of the Hamiltonian $H$, respectively. The Fermi-Dirac distribution is given by $f(x)=\left( e^{{x}/{(k_\text{B} T)}}+1\right)^{-1}$ and the bias window is defined by the electrochemical potentials $\mu_\text{L}=+eV/2$ and $\mu_\text{R}=-eV/2$ equally for the top and bottom DQD. The temperature of all leads is $T$. 
The sequential electron-tunneling regime is justified if either temperature is sufficiently high, $k_\text{B} T\gg \Gamma$, or all relevant single-electron excitation energies $\Delta E$ are sufficiently far away from the electrochemical potentials,  $\vert \Delta E{-}\mu_{\text{L/R}}\vert\gg \Gamma$~\cite{timm_2008}.
Furthermore, we checked that renormalization effects~\cite{wunsch_2005,splettstoesser_2012,stegmann_2018,stegmann_2020_2,ghoshal_2021} (similar to the Lamb shift of the energies in atoms) induced by the leads only quantitatively change our results and therefore left them out of the calculations.   

\begin{figure}[t]
  \includegraphics[width=1.\linewidth]{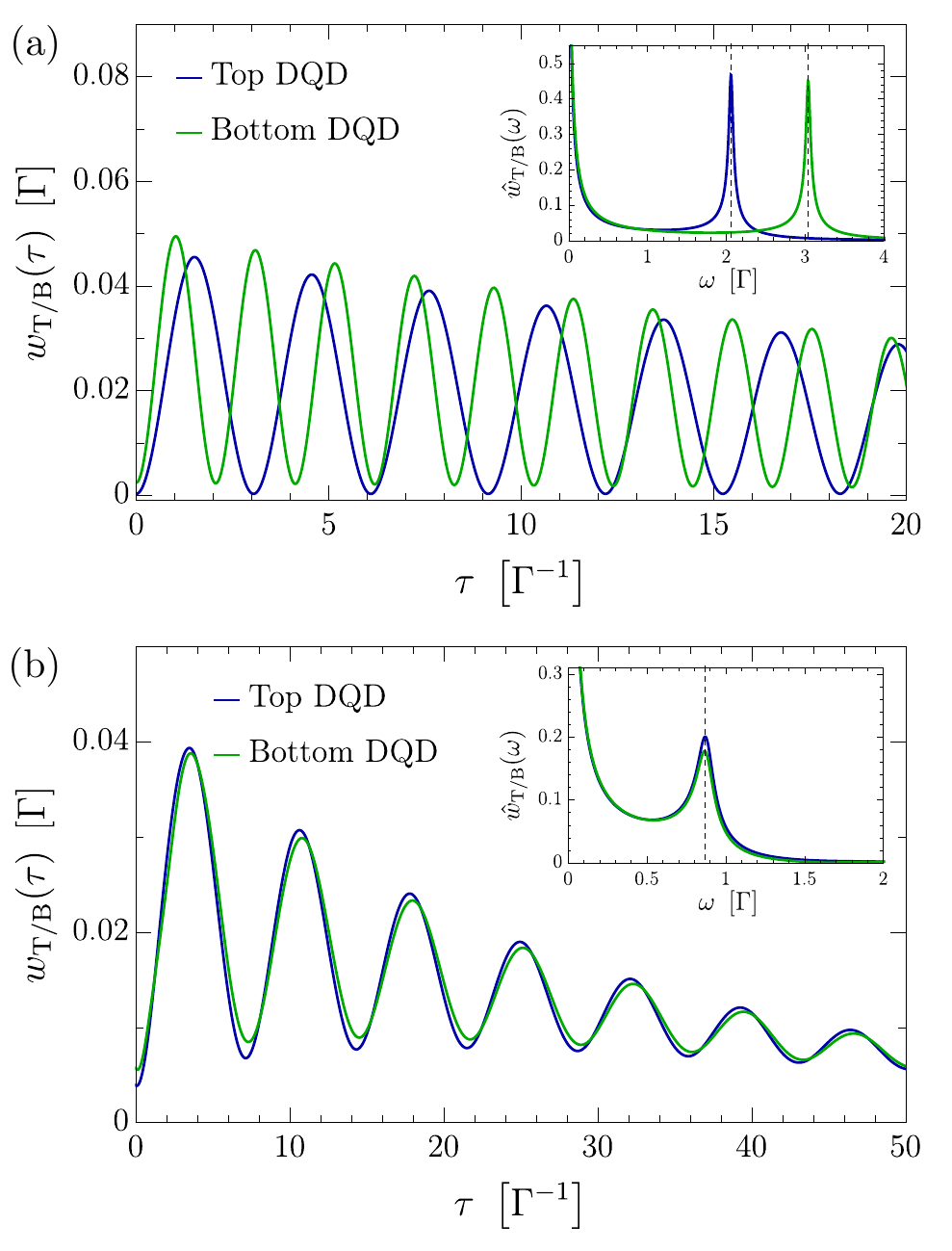}
  \caption{Synchronization. (a),(b) Waiting-time distributions $w_\text{T}(\tau)$ (blue) and $w_\text{B}(\tau)$ (green) for the top and bottom DQD, respectively. In (a), the DQDs are decoupled ($U=0\,\Gamma$), so that the detected coherent oscillations are independent of each other with distinct but different frequencies. In (b), the Coulomb interaction is $U=6\,\Gamma$, and the waiting-time distributions are almost identical $w_\text{T}(\tau)\approx w_\text{B}(\tau)$, i.e., the coherent charge oscillations are synchronized. The respective insets show the Fourier-transformed waiting-time distributions $\hat{w}_\text{T}(\omega)$ (blue) and $\hat{w}_\text{B}(\omega)$ (green) which have clear peaks at the individual frequencies $\omega_\text{T}$ and $\omega_\text{B}$ in (a) and at the common frequency $\omega_\text{S}$ in (b). The frequencies are indicated with dashed lines. The parameters are $eV=5\,\Gamma$, $k_\text{B} T=0.2\,\Gamma$, $\varepsilon=0.5\,\Gamma$, $\Delta_\text{T}=2\,\Gamma$,$\Delta_\text{B}=3\,\Gamma$, $W=25\,\Gamma$, $\Gamma_\text{L}=0.95\,\Gamma$, and $\Gamma_\text{R}=0.05\,\Gamma$.}
  \label{fig:2}
\end{figure}

\section{Synchronized Oscillations}\label{sec:synchronization}
By performing a real-time measurement of the total number of electrons $n_\alpha=n_{\alpha,\text{L}}+n_{\alpha,\text{R}}$ in either of the two DQDs, all single-electron tunneling events can be resolved as a function of time. Such a measurement does not alter the internal quantum dynamics of the system since $\left[H,n_\alpha\right]=0$, so that coherent charge oscillations are unaffected. 
A suited statistical tool to study these coherent charge oscillations is the waiting-time distribution $w_\alpha(\tau)$
which describes waiting times $\tau$ between successive tunneling-in and tunneling-out events. It can be derived via~\cite{brandes_2008}
\begin{align}\label{eq:wt}
w_\alpha(\tau)=\frac{\tr({\cal J}_{\alpha,-}e^{{\cal L}_{\alpha,0} \tau}{\cal J}_{\alpha,+}\rho_\text{st})}{\tr({\cal J}_{\alpha,+}\rho_\text{st})},
\end{align}
where we defined the jump operators according to ${\cal J}_{\alpha,s}\rho=\sum_\beta \Gamma_\beta L^{\phantom{\dagger}}_{\alpha,\beta,s} \rho L^\dagger_{\alpha,\beta,s}$ describing an electron tunneling into (${\cal J}_{\alpha,+}$) and out of  (${\cal J}_{\alpha,-}$) the $\alpha$-DQD, respectively. Between the tunneling-in and tunneling-out event, no other tunneling-in events are allowed so that the state is propagated with ${\cal L}_{\alpha,0} ={\cal L}{-}{\cal J}_{\alpha,-}$. The stationary state $\rho_\text{st}$ is defined via ${\cal L}\rho_\text{st}=0$.  The waiting-time distribution is normalized via $\int_0^\infty \mathrm{d}\tau w_\alpha(\tau)=1$.

To facilitate coherent charge oscillations, the system is tuned such that most of the time there is only one electron in each of the two DQDs. 
For this purpose, we choose $eV\ll W$, such that the charging energy $W$ is too high to allow double occupancy of a DQD. 
Moreover, by choosing $\Gamma_\text{L}\gg\Gamma_\text{R}$, one can construct a bottleneck to ensure that the electron dwells for long waiting times $\tau$ inside the DQD before tunneling out again. 
Finally, for coherent oscillations to be seen as a clear signature in the tunneling statistics we choose $\varepsilon,\Delta_{\text{T/B}} \sim\Gamma$ as well as $k_{\text{B}} T \lesssim \Gamma$. 

Within this parameter regime, only a small modification by a simple degeneracy factor $g=2$ is necessary to describe not only to spinless but also to spinful fermions (assuming double occupancy of one quantum dot is prohibited by a large onsite Coulomb interaction)~\cite{wunsch_2005}. Then, for each tunneling-in event there are $g=2$ possibilities, i.e., either a spin-up or a spin-down electron enters the system. For a tunneling-out event, however, there is only one possibility. Therefore, the Lindblad operators from Eq.~\eqref{eq:lindblad} have to be modified according to $L_{\alpha,\beta,+}\rightarrow \sqrt{g} L_{\alpha,\beta,+}$. 
However, due to the bottleneck $\Gamma_\text{L}\gg \Gamma_\text{R}$, the duration of the coherent charge oscillations is mainly limited by the tunneling-out rate, so the factor $g=2$ modifying the tunneling-in rate leads to only marginal changes in the waiting-time distribution.

In Fig.~\ref{fig:2}a, we show the waiting-time distribution $w_\text{T/B}(\tau)$ for the top (blue) and bottom (green) DQD for zero Coulomb interaction $U=0\,\Gamma$. 
We observe for both DQDs decaying oscillations with distinct but different frequencies. 
The waiting-time distribution shows clear minima indicating the times when the electron is most likely to be found in the left quantum dot, so that tunneling out of the DQD is suppressed. 
To extract the oscillation frequencies, we employ the Fourier-transformed waiting-time distribution $\hat{w}_\alpha(\omega)=\vert \int_0^\infty \mathrm{d}\tau e^{-i \omega \tau} w_\alpha(\tau)\vert$ which can be written in the form
\begin{align}\label{eq:fwt}
\hat{w}_\alpha(\omega)=\Bigg\vert \frac{\tr({\cal J}_{\alpha,-}{\left(i \omega \mathds{1}-{\cal L}_{\alpha,0} \right)^{-1}}{\cal J}_{\alpha,+}\rho_\text{st})}{\tr({\cal J}_{\alpha,+}\rho_\text{st})} \Bigg\vert.
\end{align}
The Fourier transform shows clear peaks at the frequencies $\omega_{\text{T}}=\sqrt{\Delta_\text{T}^2+\varepsilon^2}$ and  $\omega_{\text{B}}=\sqrt{\Delta_\text{B}^2+\varepsilon^2}$, cf. the insets of Fig.~\ref{fig:2}a.

By turning on the Coulomb interaction $U$ between the two DQDs, the subsystems become entangled~\cite{mansour_2020,filgueiras_2020,fanchini_2010}. In Fig.~\ref{fig:2}b, we observe that both systems show nearly identical waiting-time distributions $w_\text{T}(\tau)\approx w_\text{B}(\tau)$ and they agree on a common synchronization frequency $\omega_\text{S}<\omega_{\text{T}},\omega_{\text{B}}$ which is smaller than for the individual oscillations, cf. the insets in Fig.~\ref{fig:2}.
Thus, a simple capacitive coupling synchronizes the coherent charge oscillations in the individual DQDs into a single collective mode. 

We emphasize that the synchronized oscillations studied here have to be clearly distinguished from the effect of spontaneous quantum synchronization~\cite{roulet_2018_1,roulet_2018_2,giorgi_2013}, where initially self-sustained oscillators become synchronized.
Here, the oscillations in the DQDs are not self sustained but last only for finite waiting times $\tau$, which are stochastically distributed due to the coupling to the leads. 

\begin{figure*}[t]
  \includegraphics[width=\linewidth]{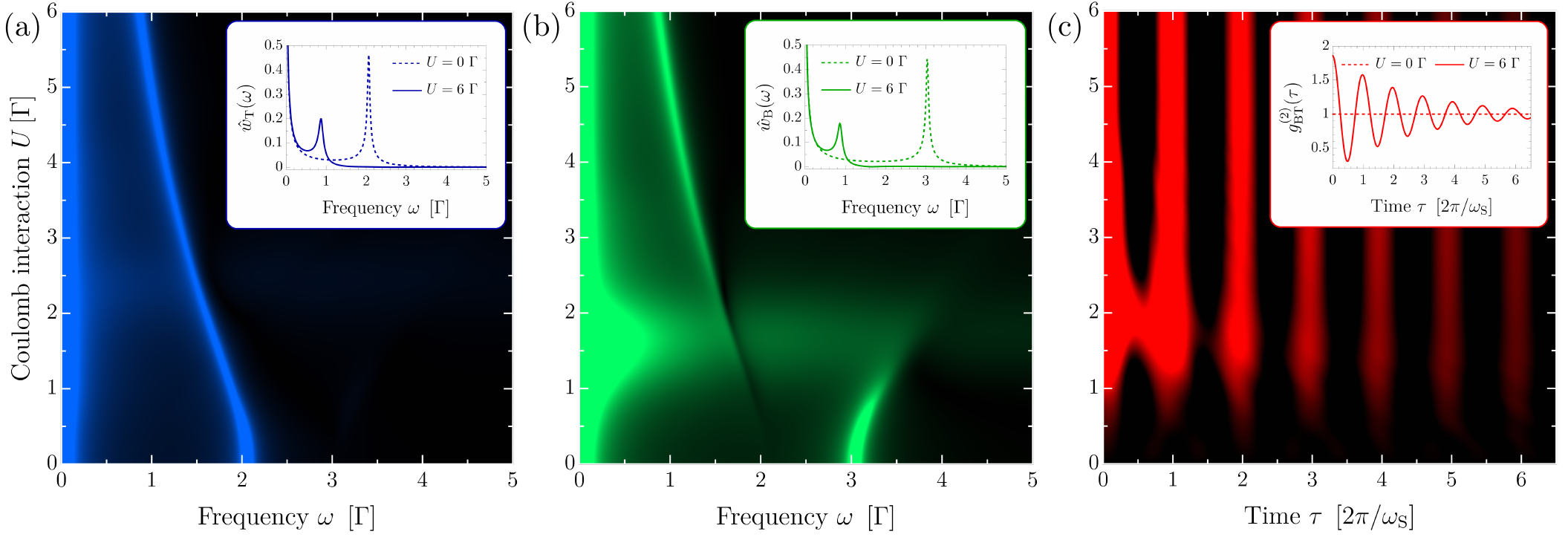}
  \caption{Frequency and Phase locking. (a),(b) Fourier-transformed waiting-time distributions $\hat{w}_\text{T}(\omega)$ and $\hat{w}_\text{B}(\omega)$ as a function of both frequency $\omega$ and Coulomb interaction $U$. In (a), the peak (blue) indicating the coherent charge oscillations of the top DQD gets gradually shifted to smaller frequencies as the Coulomb interaction $U$ increases. In (b), the peak (green) indicating the coherent charge oscillations in the bottom DQD first shifts to higher frequencies before it diminishes. As the interaction increases, a new peak appears with the same frequency as the top DQD. (c) $g^{(2)}_\text{BT}(\tau)$-correlation function as a function of both time $\tau$ and Coulomb interaction $U$ indicating temporal correlations between a tunneling-in event in the top DQD and a tunneling-out event in the bottom DQD. The maxima at  $\tau= 2 \pi n /\omega_\text{S}$ with $n=0,1,2,\ldots$ (depicted in red) suggest a phase relation of $\pi$ between the oscillations. In (a)-(c), the insets show cross sections for $U=0\,\Gamma$ and $U=6\,\Gamma$, respectively. The remaining parameters are the same as in Fig.~\ref{fig:2}.}
  \label{fig:3}
\end{figure*}

\subsection{Frequency locking}
In Fig.~\ref{fig:3}a-b, we gradually increase the Coulomb interaction $U$ between the DQDs and show the Fourier-transformed waiting-time distribution $\hat{w}_\text{T/B}(\omega)$ in (a) for the top DQD (blue) and in (b) for the bottom DQD (green).  Besides the stochastic background, we see a clear blue (green) colored signature at finite frequencies $\omega>0$ originating from the coherent charge oscillations in the top (bottom) DQD.
In Fig.~\ref{fig:3}a, the oscillations in the top DQD start at a frequency of $\omega=\omega_\text{T}\approx 2\,\Gamma$ for $U=0\,\Gamma$. Then, the frequency shifts to smaller values with increasing Coulomb interaction $U$. 
In contrast, the oscillations in the bottom DQD (see Fig.~\ref{fig:3}b) start at a frequency of about $\omega=\omega_\text{B} \approx 3\,\Gamma$ for $U=0\,\Gamma$. However, the peak gradually disappears and a new one is created as the interaction $U$ increases.
This new peak describes coherent charge oscillations in the bottom DQD where the frequency is exactly the same as in the top DQD. 
We find (see Sec.~\ref{sec:frequencies}) that the common frequency can be approximated for strong interactions by 
\begin{align}
\omega_\text{S}= \frac{\Delta_\text{T} \Delta_\text{B}}{U} + {\cal O}(1/U^2).
\end{align}
Hence, as the interaction $U$ increases, the synchronized oscillations gradually slow down until they finally disappear in the stochastic background. 
Note that in the transition region from independent to collective charge oscillations (around $U\sim eV/2$), some single-electron excitation energies $\Delta E\approx U$ of the system become resonant with the electrochemical potential $\mu_\text{L}=eV/2$, so that the condition of sequential electron tunneling is violated at low temperatures $k_\text{B} T<\Gamma$. Therefore, the detailed features in the transition region visible in Fig.~\ref{fig:3}b should be taken with a grain of salt.

\subsection{Phase locking}
To observe the phase relation of the oscillations between the top and bottom DQD, we employ the $g^{(2)}_\text{BT}(\tau)$-correlation function~\cite{emary_2012} between tunneling events. It is defined as
\begin{align}\label{eq:g2}
g^{(2)}_\text{BT}(\tau)=\frac{\tr({\cal J}_{\text{B},-}e^{{\cal L}\tau}{\cal J}_{\text{T},+}\rho_\text{st})}{\tr({\cal J}_{\text{B},-}\rho_\text{st})\tr({\cal J}_{\text{T},+}\rho_\text{st})}.
\end{align}
Thus, it measures temporal correlations between a tunneling-in event in the top DQD and a tunneling-out event in the bottom DQD. Note that in contrast to the definition of the waiting-time distribution in Eq.~\eqref{eq:wt}, here, the tunneling events are not successive and therefore the propagation between the inspected events happens with the full Liouvillian ${\cal L}$. 
In general, there is no one-to-one correspondence to the waiting-time distribution~\cite{dambach_2015}. Only for so-called renewal systems such a relation can be established~\cite{emary_2012}. 
In Fig.~\ref{fig:3}c, we show the $g^{(2)}_\text{BT}(\tau)$-correlation function as a function of both time $\tau$ and Coulomb interaction $U$. Note, that we rescaled the time variable by the period $2\pi/\omega_\text{S}$ which also depends on the interaction $U$. 
 If the Coulomb interaction is zero $U=0\,\Gamma$, the systems become disentangled and we find $g^{(2)}_\text{BT}(\tau)=1$, i.e., the statistics of the top and bottom DQD are completely uncorrelated, cf. the inset of Fig.~\ref{fig:3}c. However, for finite interaction $U>0$, we see positive and negative correlations emerging in the $g^{(2)}_\text{BT}(\tau)$ function indicating that the oscillations in the top and bottom DQD are a collective mode. In particular, the significant positive correlation (indicated in red) for times $\tau= 2 \pi n /\omega_\text{S}$ with $n=0,1,2,\ldots$ suggests that whenever an electron enters the top DQD, the probability is increased that an electron tunnels out of the bottom DQD either simultaneously or after an integer number of cycles of the oscillation. 
So if the electron in the top DQD is on the left, the electron in the bottom DQD is most likely on the right.
Thus, the observed collective charge oscillations are phase shifted by $\pi$. 

\begin{figure*}[t]
  \includegraphics[width=\linewidth]{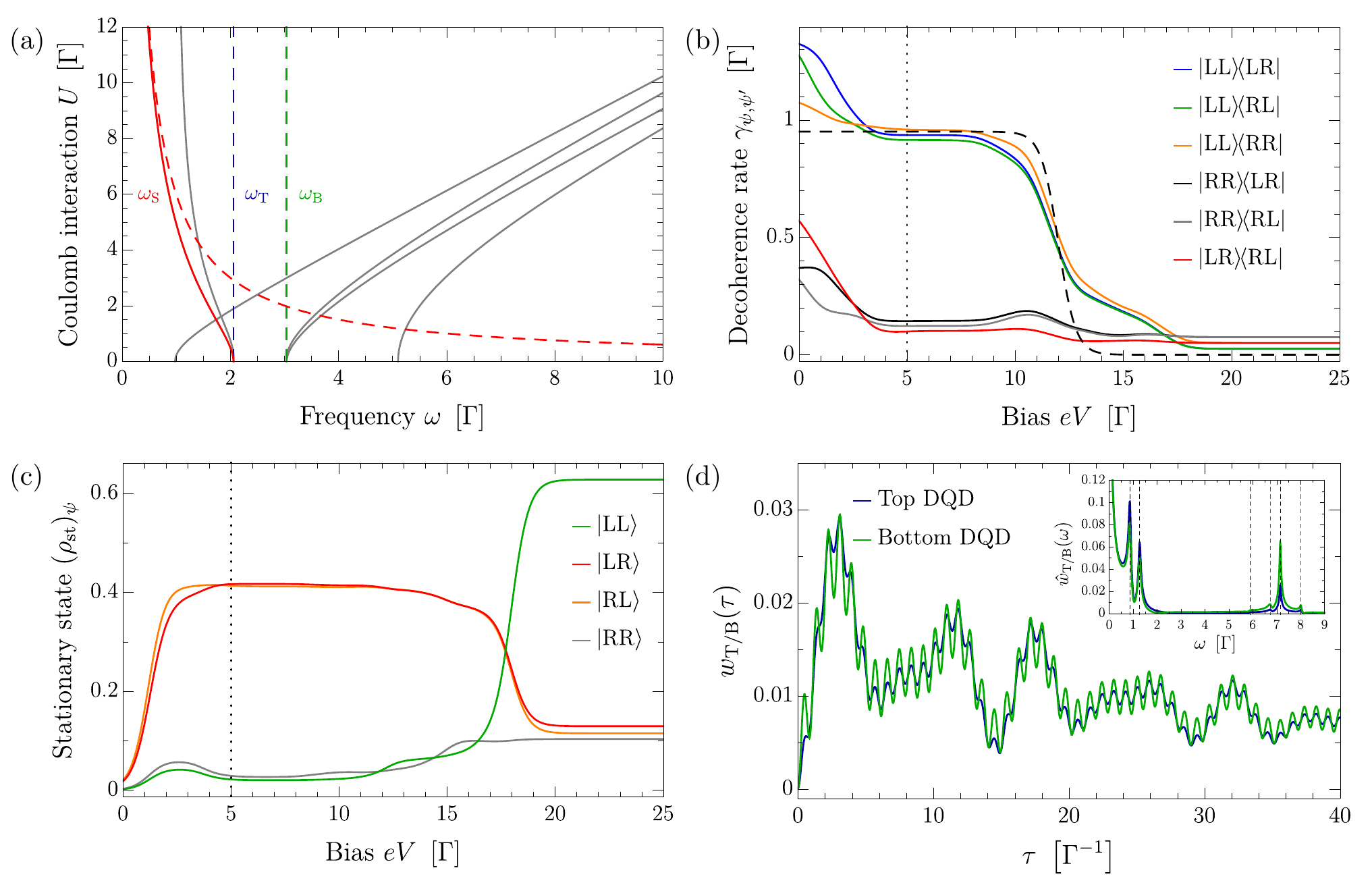}
  \caption{(a) All six oscillation frequencies (solid lines) are shown. The common synchronization frequency $\omega_\text{S}$ is depicted in red and the remaining frequencies are depicted in gray. The blue and green dashed lines indicate the individual oscillation frequencies $\omega_\text{T}$ and $\omega_\text{B}$, respectively, and the red dashed line shows the leading term of the synchronization frequency $\omega_\text{S}\approx \Delta_\text{T}\Delta_\text{B}/U$ for strong interactions $U$. (b) Decoherence rates $\gamma_{\psi\psi^\prime}$ of the coherences $\dyad{\psi}{\psi\prime}$ with  $\ket{\psi},\ket{\psi^\prime}\in\{\ket{\text{LL}}\ket{\text{LR}},\ket{\text{RL}},\ket{\text{RR}}\}$ as a function of the applied bias~$eV$. For $eV \lesssim 2U$, the coherences involving $\ket{\text{LL}}$ (blue, green and orange) decohere fast compared to the rest. Only for high bias voltages $eV\gtrsim 2U$, all six coherences decohere slowly. (c) Probabilities $(\rho_\text{st})_\psi$ to find the system in the state $\ket{\psi}$.  
For $eV\lesssim2U$, the system is most likely in either $\ket{\text{LR}}$ or $\ket{\text{RL}}$, while for $eV\gtrsim2U$,  the system is most likely in $\ket{\text{LL}}$.  
The dotted vertical line at $eV=5\,\Gamma$ in (b) and (c) indicates the bias voltage used in Fig.~\ref{fig:2} and Fig.~\ref{fig:3} to obtain the synchronized coherent charge oscillations. (d) We use $eV=20\,\Gamma$ and find that the waiting-time distributions $w_\text{T/B}(\tau)$ show beats and all six frequencies are visible in the Fourier transform $\hat{w}_\text{T/B}(\omega)$ (see inset). The remaining parameters are the same as in Fig.~\ref{fig:2}.}
  \label{fig:4}
\end{figure*}

\section{Full Set of Oscillation Frequencies}\label{sec:frequencies}
To develop a deeper understanding of the coherent charge oscillations, we study the isolated system $H$ without leads in the local basis $\ket{\beta\beta^\prime}:=\ket{\beta}_\text{T} \otimes \ket{\beta^\prime}_\text{B}$ with $\beta \in\{ 0,\text{L},\text{R},\text{D}\}$ indicating whether the DQD is empty~($0$), singly occupied with an electron in the left~$(\text{L})$ or right~$(\text{R})$ quantum dot, or doubly occupied~$(\text{D})$.
The relevant states in the electronic transport are the empty state $\ket{00}$, the singly occupied states $\{\ket{0\text{L}} ,\ket{0\text{R}} ,\ket{\text{L}0} ,\ket{\text{R}0}\}$ as well as the four doubly occupied states $\{\ket{\text{L}\text{L}} ,\ket{\text{L}\text{R}} ,\ket{\text{R}\text{L}} ,\ket{\text{R}\text{R}}\}$,  while the remaining seven states are inaccessible due to the high Coulomb interaction $W$. 
In the charge sectors spanned by $\{\ket{0\text{L}} ,\ket{0\text{R}}\}$ and $\{\ket{\text{L}0} ,\ket{\text{R}0}\}$, an electron can oscillate coherently between the left and right quantum dots at the frequency $\omega_\text{T}$ and $\omega_\text{B}$, respectively, independently of the second DQD. 
However, since $\Gamma_\text{L}\gg \Gamma_\text{R}$, the most relevant charge sector of the Hamiltonian $H$ is when the top and bottom DQD are occupied with one electron each. 
There, we can conveniently describe the degrees of freedom using the isospin operators $\vb{I}_\alpha= \boldsymbol{\sigma}^{(\alpha)}/2$, where we used the Pauli matrices $\boldsymbol{\sigma}^{(\alpha)}=\left(\sigma_x^{(\alpha)},\sigma_y^{(\alpha)},\sigma_z^{(\alpha)}\right)$ in the basis $\{ \ket{\text{L}}_\alpha,\ket{\text{R}}_\alpha\}$.
Thus, the electron being in the left or right quantum dot corresponds to the isospin up and down, respectively.
We find for the subspace spanned by $\{\ket{\text{LL}},\ket{\text{LR}},\ket{\text{RL}},\ket{\text{RR}}\}$ the following Hamiltonian 
\begin{align}\label{eq:Hred}
\tilde{H} = ~& \left(\vb{B}_\text{T} \cdot \vb{I}_\text{T}\right) \otimes \mathds{1}_\text{B}+\mathds{1}_\text{T}\otimes \left( \vb{B}_\text{B} \cdot \vb{I}_\text{B}\right) \nonumber \\
&+ 2U {I}_{\text{T},z} \otimes {I}_{\text{B},z} +\frac{U}{2}\mathds{1}_\text{T} \otimes \mathds{1}_\text{B} ,
\end{align}
where $\otimes$ denotes the tensor product. The vector $\vb{B}_\alpha=\left(-\Delta_\alpha,0,\varepsilon\right)$ takes the role of a magnetic field for the isospin and the Coulomb repulsion $U$ takes the role of a Ising-like exchange interaction between the isospins. The last term simply shifts the total energy of the system by $U/2$. For noninteracting DQDs with $U=0\,\Gamma$, both isospins $\vb{I}_\alpha$ precess independently of each other around the direction defined by $\vb{B}_\alpha$, which can be seen by the decoupled equations of motion $\dot{\vb{I}}_\alpha=\vb{B}_\alpha\times\vb{I}_\alpha $ given in the Heisenberg picture.  
Therefore, there is only one allowed frequency in each DQD which is given by $\omega_{\alpha}=\vert \vb{B}_\alpha\vert=\sqrt{\Delta_\alpha^2+\varepsilon^2}$ (indicated as a blue and green dashed line in Fig.~\ref{fig:4}a for the top and bottom DQD).

However, the dynamics of the interacting system~($U>0$) is much more complex. 
Given the four nondegenerate eigenenergies $E_\chi$ with $\chi\in\{1,2,3,4\}$ of the Hamiltonian $\tilde{H}$, one finds six distinct frequencies $E_\chi{-}E_{\chi^\prime}$ with $\chi>\chi^\prime$ (assuming $E_1{<}E_2{<}E_3{<}E_4$) which can potentially influence the coherent oscillations. In Fig.~\ref{fig:4}a, we show all six possible frequencies (solid lines) as a function of the Coulomb interaction $U$, where the synchronization frequency $\omega_\text{S}$ [cf. Fig.~\ref{fig:3}(a)] is depicted in red and the remaining frequencies are depicted in gray. To understand why two frequencies approach a constant value and four increase linearly with $U$, we examine the limit of strong interactions $U$. 
There, the eigenstates $\ket{\chi}$ of $\tilde{H}$ take a particularly simple form, namely
\begin{align}\label{eq:eigenstates}
\ket{1,2} \approx  \frac{\ket{\text{L}\text{R}}\pm\ket{\text{R}\text{L}}}{\sqrt{2}},~ \ket{3}\approx \ket{\text{R}\text{R}}, ~\ket{4}&\approx \ket{\text{L}\text{L}},
\end{align}
where corrections of order ${\cal O}(1/U)$ are neglected. The eigenstates $\ket{1}$ and $\ket{2}$ correspond to maximally entangled states, i.e., if the electron is left in one DQD, it is right in the other DQD. The respective eigenenergies are
\begin{align}
E_{1,2}\approx -\delta \mp\frac{\Delta_\text{B}\Delta_\text{T}}{2U}, \quad E_{3,4}\approx {U} +\delta \mp \varepsilon,
\end{align}
where we defined $\delta{=}\left(\Delta_\text{T}^2{+}\Delta_\text{B}^2\right)/(4U)$. Corrections of order ${\cal O}(1/U^2)$ are neglected.
Thus, in the limit of strong interactions $U$, the four frequencies $\{E_4{-}E_1,E_4{-}E_2,E_3{-}E_1,E_3{-}E_2\}$ increase linearly with $U$ and the two frequencies $\{E_4{-}E_3,E_2{-}E_1\}$ become constant, cf. Fig.~\ref{fig:4}(a).
Examining the respective eigenstates from Eq.~\eqref{eq:eigenstates}, we find that the four frequencies that increase linearly with $U$ correspond to unilateral charge oscillations, i.e., either $\ket{{\beta \text{L}}}\leftrightarrow\ket{{\beta \text{R}}}$ or $\ket{{\text{L}\beta }}\leftrightarrow\ket{{\text{R}\beta }}$ with $\beta \in \{\text{L},\text{R}\}$. The oscillations occur only in one DQD, while in the other DQD the electron sits still in either the left or the right quantum dot.
The other two frequencies, approaching a constant value,  correspond to collective charge oscillations in which the electrons oscillate either in phase, $\ket{\text{LL}}\leftrightarrow \ket{\text{RR}}$, or in antiphase, $\ket{\text{LR}}\leftrightarrow\ket{\text{RL}}$.

To elucidate the physics of the collective charge oscillations for large interactions $U$, we perform a (unitary) Schrieffer-Wolff transformation ${H^\prime}=e^S{\tilde{H}}e^{-S}$ to effectively decouple low-energy states ($\ket{\text{LR}},\ket{\text{RL}}$) from high-energy states ($\ket{\text{LL}},\ket{\text{RR}}$). Therefore, we artificially decompose the Hamiltonian $\tilde{H}=H_0+ V$ into a diagonal part $H_0$ and an off-diagonal perturbation $V\propto \Delta_\alpha$, where $\Delta_\alpha \ll U$. By choosing the antihermitian generator $S$ such that $\left[ H_0, S \right] =V$, we find $H^\prime=H_0+\left[ S, V \right]/2+{\cal O}(V^3)$, i.e., the linear order in $V$ has been eliminated. The remaining degrees of freedom of $S$ have been chosen such that $H^\prime$ becomes blockdiagonal 
\begin{align}\label{eq:schriefferwolff}
H^\prime= \lbrack (U{+}\delta) \mathds{1}_\text{IP}+\vb{B}_\text{IP} \cdot \vb{I}_\text{IP}\rbrack \oplus \lbrack-\delta \mathds{1}_\text{AP}+\vb{B}_\text{AP} \cdot \vb{I}_\text{AP}\rbrack 
\end{align}
in the basis $\{\ket{\text{LL}},\ket{\text{RR}},\ket{\text{LR}},\ket{\text{RL}}\}$, where corrections of order ${\cal O}(1/U^2)$ are neglected. 
Thus, the Hamiltonian decouples into a direct sum $H^\prime=H_\text{IP}\oplus H_\text{AP}$.
The effective high-energy Hamiltonian $H_\text{IP}$, which is linear in $U$, describes in-phase oscillations ($\text{IP}$) in the subspace $\{ \ket{\text{LL}},\ket{\text{RR}}\}$, while the effective low-energy Hamiltonian $H_\text{AP}$, which is of order $1/U$, describes antiphase oscillations ($\text{AP}$) in the subspace $\{ \ket{\text{LR}},\ket{\text{RL}}\}$.
Analogously to Eq.~\eqref{eq:Hred}, the isospins are defined via $\vb{I}_{\nu}=  \boldsymbol{\sigma}^{(\nu)}/2$, where now the Pauli matrices are given in the basis $\{\ket{\text{LL}},\ket{\text{RR}}\}$ and $\{\ket{\text{LR}},\ket{\text{RL}}\}$ for $\nu=\text{IP}$ and $\nu=\text{AP}$, respectively. Furthermore, the effective magnetic fields for the isospins are given by $\vb{B}_\text{IP}=\left(\Delta_\text{T}\Delta_{\text{B}}/U,0,2\varepsilon\right)$ and $\vb{B}_\text{AP}=\left(-\Delta_\text{T}\Delta_{\text{B}}/U,0,0\right)$. 
They give rise to collective in-phase oscillations, $\dot{\vb{I}}_\text{IP}=\vb{B}_\text{IP}\times \vb{I}_\text{IP}$, with frequency $\omega_{\text{IP}}=\vert \vb{B}_\text{IP}\vert= 2\varepsilon+{\cal O}(1/U^2)$ and collective antiphase oscillations, $\dot{\vb{I}}_\text{AP}=\vb{B}_\text{AP}\times \vb{I}_\text{AP}$, with frequency $\omega_{\text{AP}}=\vert \vb{B}_\text{AP}\vert=\Delta_\text{T}\Delta_\text{B}/U$. 
Note that the decoupled subspaces are highly entangled, since no pure state $\ket{\psi}_{\text{IP}}=\cos(\theta/2)\ket{\text{LL}} +e^{i \phi} \sin(\theta/2)\ket{\text{RR}}$ or  $\ket{\psi}_{\text{AP}}=\cos(\theta/2)\ket{\text{LR}} +e^{i \phi} \sin(\theta/2)\ket{\text{RL}}$ on the respective Bloch sphere can be written as a product state $\ket{\psi}_\text{T}\otimes \ket{\psi}_\text{B}$, except for the poles ($\theta=0$ and $\theta=\pi$). Therefore, the entanglement arises quite naturally in the dynamics.

We emphasize that in the non-equilibrium situation of electron transport, not only the states with the lowest energy contribute, but all states where each DQD is either empty or singly occupied.
Nonetheless, we find that for a specific bias voltage $eV$, out of all six possible frequencies only one is singled out, namely the one describing collective antiphase oscillations (cf. Fig.~\ref{fig:3}) with approximate frequency $\omega_\text{S}\approx \omega_\text{AP}$ [red dashed line in Fig.~\ref{fig:4}(a)]. 

\section{Environment-induced Decoherence}\label{sec:decoherence}
To find out why only one out of six modes is visible in the coherent charge oscillations when the DQDs are coupled to the environment, we inspect the decoherence induced by the leads. 
Other decoherence mechanisms (e.g. due to coupling to phonons~\cite{brandes_2008}), are not considered here.
We define the decoherence rates as
\begin{align}
\gamma_{\psi ,\psi^\prime}=-\text{Re}\left[\tr( \dyad{{\psi}}{{\psi^\prime}}^\dagger {\cal L}  \dyad{{\psi}}{{\psi^\prime}})\right]
\end{align}
for the six relevant coherences  $\dyad{{\psi}}{{\psi^\prime}}$ with $\ket{\psi},\ket{\psi^\prime}\in\{\ket{\text{LL}},\ket{\text{LR}},\ket{\text{RL}},\ket{\text{RR}}\}$ and $\psi\neq\psi^\prime$ as a function of the applied bias $eV$, see Fig.~\ref{fig:4}b.

In the regime where synchronization can be observed (dotted line at $eV=5\,\Gamma$), we see that the decoherence rates $\gamma_{\text{LL},\text{LR}},\gamma_{\text{LL},\text{RL}}$ and $\gamma_{\text{LL},\text{RR}}$ (blue, green and orange) are much larger than the remaining decoherence rates $\gamma_{\text{RR},\text{LR}},\gamma_{\text{RR},\text{RL}}$ and $\gamma_{\text{LR},\text{RL}}$ (black, gray and red). Thus, all coherences involving the high-energy state $\ket{\text{LL}}$ decohere fast, so the number of relevant coherences is reduced from six to three. 
This effect can be understood by studying the lead-induced decoherence mechanism, which has its origin in virtual charge fluctuations into and out of the metallic leads.
Since $\Gamma_\text{L}\gg\Gamma_\text{R}$, this decoherence effect is mainly caused by the left leads. 
Furthermore, since the tunneling-in event of a second electron into the DQD is suppressed by a large Coulomb repulsion $W\gg eV$, we have to consider only charge fluctuations where an electron first virtually tunnels out of the system and then tunnels in again. 
For the state $\ket{\text{R}\text{R}}$, such a tunneling-out event at the left leads is trivially suppressed by the geometry of the setup.
Also for the state $\ket{\text{L}\text{R}}$ ($\ket{\text{R}\text{L}}$), the tunneling-out event into the top (bottom) left lead is ineffective because the electron has to virtually tunnel against the natural direction of the applied bias $eV$. Since there are no free states available well below the left Fermi energy $\mu_\text{L}=eV/2$, the necessary tunneling-out events are strongly suppressed.
In contrast, for coherences involving the high-energy state $\ket{\text{L}\text{L}}$, the additional charging energy $U$ supplied by the coupling between the DQDs can be large enough to overcome the barrier set by $\mu_\text{L}=eV/2$ and tunneling against the natural direction becomes possible. 
As a rough approximation, we find the following form for the decoherence rates (black dashed line in Fig.~\ref{fig:4}b)
\begin{align}
\gamma_{\text{LL},\psi^\prime}\approx \Gamma_\text{L} \left[ 1 - f\left(U-\frac{eV}{2}\right)\right],
\end{align}
where we have neglected contributions of $\Delta_{\text{T},\text{B}}$ and $\varepsilon$ to the excitation energies. Hence, all coherences involving the state $\ket{\text{L}\text{L}}$ can effectively decohere if the applied bias fulfills $eV\lesssim 2U$. 

Although the decoherence rates  $\gamma_{\text{RR},\text{LR}}$ and $\gamma_{\text{RR},\text{RL}}$ are small, it is still unlikely to find the system in the state $\ket{\text{RR}}$. Therefore, also coherences involving the high-energy state $\ket{\text{RR}}$ are of minor importance for the coherent charge oscillations. In Fig~\ref{fig:4}c, we show the probability $(\rho_\text{st})_{\psi}=\mel{\psi}{\rho_\text{st}}{\psi}$ to find the system in any of the configurations $\ket{\psi}\in\{\ket{\text{LL}},\ket{\text{LR}},\ket{\text{RL}},\ket{\text{RR}}\}$ as a function of the applied bias $eV$.  For the regime where synchronization can be observed (dotted line at $eV=5\,\Gamma$), there is a high probability of finding the system in $\ket{\text{LR}}$ (red) or $\ket{\text{RL}}$ (orange), while it is rather unlikely to find it in $\ket{\text{LL}}$ (green) or $\ket{\text{RR}}$ (gray).  With increasing bias $eV$ the probability to be in $\ket{\text{LL}}$ increases drastically, because the decay mechanism described above becomes more and more ineffective. For the state $\ket{\text{RR}}$, however, the probability remains small more or less independent of the applied bias $eV$. This can again be explained by the geometry of the setup, since electrons tunnel into the system only from the left side, so that a direct transition to $\ket{\text{RR}}$ from any state with only one electron is suppressed. 
Thus, the dominant coherences are expected to appear only between the states $\ket{\text{LR}}$ and $\ket{\text{RL}}$. This finally explains why only the collective antiphase coherent charge oscillations with frequency $\omega_\text{S}$ are visible in the electron statistics. 

If, however, we increase the applied bias to $eV\gtrsim 2U$, the decoherence mechanism becomes so ineffective that the waiting-time distribution $w_\alpha(\tau)$ shows beats, see Fig.~\ref{fig:4}d. Then, all six possible frequencies become visible in the electron transport when analyzing the Fourier-transformed waiting-time distributions $\hat{w}_\alpha(\omega)$, see the inset of Fig.~\ref{fig:4}d.  

\section{Conclusions}\label{sec:conclusion}
We studied coherent charge oscillations in the electron transport through two serial double quantum dots when they are coupled to each other via Coulomb interaction. 
Whereas for zero interaction the charge oscillations in the double quantum dots are independent of each other with distinct but different frequencies, we find that as the interaction increases the individual oscillations become synchronized with a common frequency and a fixed phase relation of $\pi$. 
Although the entangled system has potentially six frequencies that may occur in the coherent charge oscillations, we find that an appropriately chosen bias voltage leads to the preference of only one frequency that is visible in the electron transport by means of waiting-time distributions. The suppression of all remaining frequencies is a non-equilibrium effect where the charging energy supplied by the Coulomb interaction is utilized to tunnel against the natural bias direction. This enables an effective decay of all but one oscillation mode, where the electrons in the top and bottom double quantum dot collectively oscillate in antiphase. 

\section*{Acknowledgements}
We gratefully acknowledge funding by the Deutsche Forschungs\-gemeinschaft (DFG, German Research Foundation) -- Project 278162697 -- SFB 1242.
PS acknowledges support from the German National Academy of Sciences Leopoldina (Grant No. LPDS 2019-10).


\begin{thebibliography}{54}%
\makeatletter
\providecommand \@ifxundefined [1]{%
 \@ifx{#1\undefined}
}%
\providecommand \@ifnum [1]{%
 \ifnum #1\expandafter \@firstoftwo
 \else \expandafter \@secondoftwo
 \fi
}%
\providecommand \@ifx [1]{%
 \ifx #1\expandafter \@firstoftwo
 \else \expandafter \@secondoftwo
 \fi
}%
\providecommand \natexlab [1]{#1}%
\providecommand \enquote  [1]{``#1''}%
\providecommand \bibnamefont  [1]{#1}%
\providecommand \bibfnamefont [1]{#1}%
\providecommand \citenamefont [1]{#1}%
\providecommand \href@noop [0]{\@secondoftwo}%
\providecommand \href [0]{\begingroup \@sanitize@url \@href}%
\providecommand \@href[1]{\@@startlink{#1}\@@href}%
\providecommand \@@href[1]{\endgroup#1\@@endlink}%
\providecommand \@sanitize@url [0]{\catcode `\\12\catcode `\$12\catcode
  `\&12\catcode `\#12\catcode `\^12\catcode `\_12\catcode `\%12\relax}%
\providecommand \@@startlink[1]{}%
\providecommand \@@endlink[0]{}%
\providecommand \url  [0]{\begingroup\@sanitize@url \@url }%
\providecommand \@url [1]{\endgroup\@href {#1}{\urlprefix }}%
\providecommand \urlprefix  [0]{URL }%
\providecommand \Eprint [0]{\href }%
\providecommand \doibase [0]{https://doi.org/}%
\providecommand \selectlanguage [0]{\@gobble}%
\providecommand \bibinfo  [0]{\@secondoftwo}%
\providecommand \bibfield  [0]{\@secondoftwo}%
\providecommand \translation [1]{[#1]}%
\providecommand \BibitemOpen [0]{}%
\providecommand \bibitemStop [0]{}%
\providecommand \bibitemNoStop [0]{.\EOS\space}%
\providecommand \EOS [0]{\spacefactor3000\relax}%
\providecommand \BibitemShut  [1]{\csname bibitem#1\endcsname}%
\let\auto@bib@innerbib\@empty
\bibitem [{\citenamefont {Zagoskin}(2011)}]{zagoskin_2011}%
  \BibitemOpen
  \bibfield  {author} {\bibinfo {author} {\bibfnamefont {A.~M.}\ \bibnamefont
  {Zagoskin}},\ }\href@noop {} {\emph {\bibinfo {title} {Quantum engineering:
  theory and design of quantum coherent structures}}}\ (\bibinfo  {publisher}
  {Cambridge University Press},\ \bibinfo {year} {2011})\BibitemShut {NoStop}%
\bibitem [{\citenamefont {Takakura}\ \emph {et~al.}(2014)\citenamefont
  {Takakura}, \citenamefont {Noiri}, \citenamefont {Obata}, \citenamefont
  {Otsuka}, \citenamefont {Yoneda}, \citenamefont {Yoshida},\ and\
  \citenamefont {Tarucha}}]{takatura_2014}%
  \BibitemOpen
  \bibfield  {author} {\bibinfo {author} {\bibfnamefont {T.}~\bibnamefont
  {Takakura}}, \bibinfo {author} {\bibfnamefont {A.}~\bibnamefont {Noiri}},
  \bibinfo {author} {\bibfnamefont {T.}~\bibnamefont {Obata}}, \bibinfo
  {author} {\bibfnamefont {T.}~\bibnamefont {Otsuka}}, \bibinfo {author}
  {\bibfnamefont {J.}~\bibnamefont {Yoneda}}, \bibinfo {author} {\bibfnamefont
  {K.}~\bibnamefont {Yoshida}},\ and\ \bibinfo {author} {\bibfnamefont
  {S.}~\bibnamefont {Tarucha}},\ }\bibfield  {title} {\bibinfo {title} {Single
  to quadruple quantum dots with tunable tunnel couplings},\ }\href
  {https://doi.org/10.1063/1.4869108} {\bibfield  {journal} {\bibinfo
  {journal} {Appl. Phys. Lett.}\ }\textbf {\bibinfo {volume} {104}},\ \bibinfo
  {pages} {113109} (\bibinfo {year} {2014})}\BibitemShut {NoStop}%
\bibitem [{\citenamefont {Mukhopadhyay}\ \emph {et~al.}(2018)\citenamefont
  {Mukhopadhyay}, \citenamefont {Dehollain}, \citenamefont {Reichl},
  \citenamefont {Wegscheider},\ and\ \citenamefont {Vandersypen}}]{mukho_2018}%
  \BibitemOpen
  \bibfield  {author} {\bibinfo {author} {\bibfnamefont {U.}~\bibnamefont
  {Mukhopadhyay}}, \bibinfo {author} {\bibfnamefont {J.~P.}\ \bibnamefont
  {Dehollain}}, \bibinfo {author} {\bibfnamefont {C.}~\bibnamefont {Reichl}},
  \bibinfo {author} {\bibfnamefont {W.}~\bibnamefont {Wegscheider}},\ and\
  \bibinfo {author} {\bibfnamefont {L.~M.~K.}\ \bibnamefont {Vandersypen}},\
  }\bibfield  {title} {\bibinfo {title} {A 2 x 2 quantum dot array with
  controllable inter-dot tunnel couplings},\ }\href
  {https://doi.org/10.1063/1.5025928} {\bibfield  {journal} {\bibinfo
  {journal} {Appl. Phys. Lett.}\ }\textbf {\bibinfo {volume} {112}},\ \bibinfo
  {pages} {183505} (\bibinfo {year} {2018})}\BibitemShut {NoStop}%
\bibitem [{\citenamefont {Banszerus}\ \emph {et~al.}(2020)\citenamefont
  {Banszerus}, \citenamefont {M{\"o}ller}, \citenamefont {Icking},
  \citenamefont {Watanabe}, \citenamefont {Taniguchi}, \citenamefont {Volk},\
  and\ \citenamefont {Stampfer}}]{banszerus_2020}%
  \BibitemOpen
  \bibfield  {author} {\bibinfo {author} {\bibfnamefont {L.}~\bibnamefont
  {Banszerus}}, \bibinfo {author} {\bibfnamefont {S.}~\bibnamefont
  {M{\"o}ller}}, \bibinfo {author} {\bibfnamefont {E.}~\bibnamefont {Icking}},
  \bibinfo {author} {\bibfnamefont {K.}~\bibnamefont {Watanabe}}, \bibinfo
  {author} {\bibfnamefont {T.}~\bibnamefont {Taniguchi}}, \bibinfo {author}
  {\bibfnamefont {C.}~\bibnamefont {Volk}},\ and\ \bibinfo {author}
  {\bibfnamefont {C.}~\bibnamefont {Stampfer}},\ }\bibfield  {title} {\bibinfo
  {title} {Single-electron double quantum dots in bilayer graphene},\ }\href
  {https://doi.org/10.1021/acs.nanolett.9b05295} {\bibfield  {journal}
  {\bibinfo  {journal} {Nano Lett.}\ }\textbf {\bibinfo {volume} {20}},\
  \bibinfo {pages} {2005} (\bibinfo {year} {2020})}\BibitemShut {NoStop}%
\bibitem [{\citenamefont {Brandes}(2008)}]{brandes_2008}%
  \BibitemOpen
  \bibfield  {author} {\bibinfo {author} {\bibfnamefont {T.}~\bibnamefont
  {Brandes}},\ }\bibfield  {title} {\bibinfo {title} {Waiting times and noise
  in single particle transport},\ }\href
  {https://doi.org/https://doi.org/10.1002/andp.200810306} {\bibfield
  {journal} {\bibinfo  {journal} {Ann. Phys. (Berl.)}\ }\textbf {\bibinfo
  {volume} {17}},\ \bibinfo {pages} {477} (\bibinfo {year} {2008})}\BibitemShut
  {NoStop}%
\bibitem [{\citenamefont {Kambly}\ and\ \citenamefont
  {Flindt}(2013)}]{kambly_2013}%
  \BibitemOpen
  \bibfield  {author} {\bibinfo {author} {\bibfnamefont {D.}~\bibnamefont
  {Kambly}}\ and\ \bibinfo {author} {\bibfnamefont {C.}~\bibnamefont
  {Flindt}},\ }\bibfield  {title} {\bibinfo {title} {Time-dependent factorial
  cumulants in interacting nano-scale systems},\ }\href
  {https://doi.org/10.1007/s10825-013-0464-9} {\bibfield  {journal} {\bibinfo
  {journal} {J. Comput. Electron.}\ }\textbf {\bibinfo {volume} {12}},\
  \bibinfo {pages} {331} (\bibinfo {year} {2013})}\BibitemShut {NoStop}%
\bibitem [{\citenamefont {Korotkov}\ and\ \citenamefont
  {Averin}(2001)}]{korotkov_2001}%
  \BibitemOpen
  \bibfield  {author} {\bibinfo {author} {\bibfnamefont {A.~N.}\ \bibnamefont
  {Korotkov}}\ and\ \bibinfo {author} {\bibfnamefont {D.~V.}\ \bibnamefont
  {Averin}},\ }\bibfield  {title} {\bibinfo {title} {Continuous weak
  measurement of quantum coherent oscillations},\ }\href
  {https://doi.org/10.1103/PhysRevB.64.165310} {\bibfield  {journal} {\bibinfo
  {journal} {Phys. Rev. B}\ }\textbf {\bibinfo {volume} {64}},\ \bibinfo
  {pages} {165310} (\bibinfo {year} {2001})}\BibitemShut {NoStop}%
\bibitem [{\citenamefont {Korotkov}(1999)}]{korotkov_1999}%
  \BibitemOpen
  \bibfield  {author} {\bibinfo {author} {\bibfnamefont {A.~N.}\ \bibnamefont
  {Korotkov}},\ }\bibfield  {title} {\bibinfo {title} {Continuous quantum
  measurement of a double dot},\ }\href
  {https://doi.org/10.1103/PhysRevB.60.5737} {\bibfield  {journal} {\bibinfo
  {journal} {Phys. Rev. B}\ }\textbf {\bibinfo {volume} {60}},\ \bibinfo
  {pages} {5737} (\bibinfo {year} {1999})}\BibitemShut {NoStop}%
\bibitem [{\citenamefont {Gurvitz}(1997)}]{gurvitz_1997}%
  \BibitemOpen
  \bibfield  {author} {\bibinfo {author} {\bibfnamefont {S.~A.}\ \bibnamefont
  {Gurvitz}},\ }\bibfield  {title} {\bibinfo {title} {Measurements with a
  noninvasive detector and dephasing mechanism},\ }\href
  {https://doi.org/10.1103/PhysRevB.56.15215} {\bibfield  {journal} {\bibinfo
  {journal} {Phys. Rev. B}\ }\textbf {\bibinfo {volume} {56}},\ \bibinfo
  {pages} {15215} (\bibinfo {year} {1997})}\BibitemShut {NoStop}%
\bibitem [{\citenamefont {Kushihara}\ \emph {et~al.}(2012)\citenamefont
  {Kushihara}, \citenamefont {Okuyama},\ and\ \citenamefont
  {Eto}}]{kushihara_2012}%
  \BibitemOpen
  \bibfield  {author} {\bibinfo {author} {\bibfnamefont {J.}~\bibnamefont
  {Kushihara}}, \bibinfo {author} {\bibfnamefont {R.}~\bibnamefont {Okuyama}},\
  and\ \bibinfo {author} {\bibfnamefont {M.}~\bibnamefont {Eto}},\ }\bibfield
  {title} {\bibinfo {title} {Coherent and incoherent current drag in coupled
  quantum dots},\ }\href {https://doi.org/10.1088/1742-6596/400/4/042037}
  {\bibfield  {journal} {\bibinfo  {journal} {J. Phys. Conf. Ser.}\ }\textbf
  {\bibinfo {volume} {400}},\ \bibinfo {pages} {042037} (\bibinfo {year}
  {2012})}\BibitemShut {NoStop}%
\bibitem [{\citenamefont {Fujisawa}\ \emph {et~al.}(2011)\citenamefont
  {Fujisawa}, \citenamefont {Shinkai}, \citenamefont {Hayashi},\ and\
  \citenamefont {Ota}}]{fujisawa_2011}%
  \BibitemOpen
  \bibfield  {author} {\bibinfo {author} {\bibfnamefont {T.}~\bibnamefont
  {Fujisawa}}, \bibinfo {author} {\bibfnamefont {G.}~\bibnamefont {Shinkai}},
  \bibinfo {author} {\bibfnamefont {T.}~\bibnamefont {Hayashi}},\ and\ \bibinfo
  {author} {\bibfnamefont {T.}~\bibnamefont {Ota}},\ }\bibfield  {title}
  {\bibinfo {title} {Multiple two-qubit operations for a coupled semiconductor
  charge qubit},\ }\href
  {https://doi.org/https://doi.org/10.1016/j.physe.2010.07.040} {\bibfield
  {journal} {\bibinfo  {journal} {Phys. E: Low-Dimens. Syst. Nanostructures.}\
  }\textbf {\bibinfo {volume} {43}},\ \bibinfo {pages} {730} (\bibinfo {year}
  {2011})}\BibitemShut {NoStop}%
\bibitem [{\citenamefont {Mansour}\ \emph {et~al.}(2020)\citenamefont
  {Mansour}, \citenamefont {Siyouri}, \citenamefont {Faqir},\ and\
  \citenamefont {Baz}}]{mansour_2020}%
  \BibitemOpen
  \bibfield  {author} {\bibinfo {author} {\bibfnamefont {H.~A.}\ \bibnamefont
  {Mansour}}, \bibinfo {author} {\bibfnamefont {F.-Z.}\ \bibnamefont
  {Siyouri}}, \bibinfo {author} {\bibfnamefont {M.}~\bibnamefont {Faqir}},\
  and\ \bibinfo {author} {\bibfnamefont {M.~E.}\ \bibnamefont {Baz}},\
  }\bibfield  {title} {\bibinfo {title} {Quantum correlations dynamics in two
  coupled semiconductor {InAs} quantum dots},\ }\href
  {https://doi.org/10.1088/1402-4896/aba666} {\bibfield  {journal} {\bibinfo
  {journal} {Phys. Scr.}\ }\textbf {\bibinfo {volume} {95}},\ \bibinfo {pages}
  {095101} (\bibinfo {year} {2020})}\BibitemShut {NoStop}%
\bibitem [{\citenamefont {Filgueiras}\ \emph {et~al.}(2020)\citenamefont
  {Filgueiras}, \citenamefont {Rojas},\ and\ \citenamefont
  {Rojas}}]{filgueiras_2020}%
  \BibitemOpen
  \bibfield  {author} {\bibinfo {author} {\bibfnamefont {C.}~\bibnamefont
  {Filgueiras}}, \bibinfo {author} {\bibfnamefont {O.}~\bibnamefont {Rojas}},\
  and\ \bibinfo {author} {\bibfnamefont {M.}~\bibnamefont {Rojas}},\ }\bibfield
   {title} {\bibinfo {title} {Thermal entanglement and correlated coherence in
  two coupled double quantum dots systems},\ }\href
  {https://doi.org/https://doi.org/10.1002/andp.202000207} {\bibfield
  {journal} {\bibinfo  {journal} {Ann. Phys. (Berl.)}\ }\textbf {\bibinfo
  {volume} {532}},\ \bibinfo {pages} {2000207} (\bibinfo {year}
  {2020})}\BibitemShut {NoStop}%
\bibitem [{\citenamefont {Fanchini}\ \emph {et~al.}(2010)\citenamefont
  {Fanchini}, \citenamefont {Castelano},\ and\ \citenamefont
  {Caldeira}}]{fanchini_2010}%
  \BibitemOpen
  \bibfield  {author} {\bibinfo {author} {\bibfnamefont {F.~F.}\ \bibnamefont
  {Fanchini}}, \bibinfo {author} {\bibfnamefont {L.~K.}\ \bibnamefont
  {Castelano}},\ and\ \bibinfo {author} {\bibfnamefont {A.~O.}\ \bibnamefont
  {Caldeira}},\ }\bibfield  {title} {\bibinfo {title} {Entanglement versus
  quantum discord in two coupled double quantum dots},\ }\href
  {https://doi.org/10.1088/1367-2630/12/7/073009} {\bibfield  {journal}
  {\bibinfo  {journal} {New J. Phys.}\ }\textbf {\bibinfo {volume} {12}},\
  \bibinfo {pages} {073009} (\bibinfo {year} {2010})}\BibitemShut {NoStop}%
\bibitem [{\citenamefont {Kim}\ \emph {et~al.}(2015)\citenamefont {Kim},
  \citenamefont {Ward}, \citenamefont {Simmons}, \citenamefont {Gamble},
  \citenamefont {Blume-Kohout}, \citenamefont {Nielsen}, \citenamefont
  {Savage}, \citenamefont {Lagally}, \citenamefont {Friesen}, \citenamefont
  {Coppersmith},\ and\ \citenamefont {Eriksson}}]{kim_2015}%
  \BibitemOpen
  \bibfield  {author} {\bibinfo {author} {\bibfnamefont {D.}~\bibnamefont
  {Kim}}, \bibinfo {author} {\bibfnamefont {D.~R.}\ \bibnamefont {Ward}},
  \bibinfo {author} {\bibfnamefont {C.~B.}\ \bibnamefont {Simmons}}, \bibinfo
  {author} {\bibfnamefont {J.~K.}\ \bibnamefont {Gamble}}, \bibinfo {author}
  {\bibfnamefont {R.}~\bibnamefont {Blume-Kohout}}, \bibinfo {author}
  {\bibfnamefont {E.}~\bibnamefont {Nielsen}}, \bibinfo {author} {\bibfnamefont
  {D.~E.}\ \bibnamefont {Savage}}, \bibinfo {author} {\bibfnamefont {M.~G.}\
  \bibnamefont {Lagally}}, \bibinfo {author} {\bibfnamefont {M.}~\bibnamefont
  {Friesen}}, \bibinfo {author} {\bibfnamefont {S.~N.}\ \bibnamefont
  {Coppersmith}},\ and\ \bibinfo {author} {\bibfnamefont {M.~A.}\ \bibnamefont
  {Eriksson}},\ }\bibfield  {title} {\bibinfo {title} {Microwave-driven
  coherent operation of a semiconductor quantum dot charge qubit},\ }\href
  {https://doi.org/10.1038/nnano.2014.336} {\bibfield  {journal} {\bibinfo
  {journal} {Nat. Nanotechnol.}\ }\textbf {\bibinfo {volume} {10}},\ \bibinfo
  {pages} {243} (\bibinfo {year} {2015})}\BibitemShut {NoStop}%
\bibitem [{\citenamefont {Srinivasa}\ \emph {et~al.}(2013)\citenamefont
  {Srinivasa}, \citenamefont {Nowack}, \citenamefont {Shafiei}, \citenamefont
  {Vandersypen},\ and\ \citenamefont {Taylor}}]{srinivasa_2013}%
  \BibitemOpen
  \bibfield  {author} {\bibinfo {author} {\bibfnamefont {V.}~\bibnamefont
  {Srinivasa}}, \bibinfo {author} {\bibfnamefont {K.~C.}\ \bibnamefont
  {Nowack}}, \bibinfo {author} {\bibfnamefont {M.}~\bibnamefont {Shafiei}},
  \bibinfo {author} {\bibfnamefont {L.~M.~K.}\ \bibnamefont {Vandersypen}},\
  and\ \bibinfo {author} {\bibfnamefont {J.~M.}\ \bibnamefont {Taylor}},\
  }\bibfield  {title} {\bibinfo {title} {Simultaneous spin-charge relaxation in
  double quantum dots},\ }\href
  {https://doi.org/10.1103/PhysRevLett.110.196803} {\bibfield  {journal}
  {\bibinfo  {journal} {Phys. Rev. Lett.}\ }\textbf {\bibinfo {volume} {110}},\
  \bibinfo {pages} {196803} (\bibinfo {year} {2013})}\BibitemShut {NoStop}%
\bibitem [{\citenamefont {Cao}\ \emph {et~al.}(2013)\citenamefont {Cao},
  \citenamefont {Li}, \citenamefont {Tu}, \citenamefont {Wang}, \citenamefont
  {Zhou}, \citenamefont {Xiao}, \citenamefont {Guo}, \citenamefont {Jiang},\
  and\ \citenamefont {Guo}}]{cao_2013}%
  \BibitemOpen
  \bibfield  {author} {\bibinfo {author} {\bibfnamefont {G.}~\bibnamefont
  {Cao}}, \bibinfo {author} {\bibfnamefont {H.-O.}\ \bibnamefont {Li}},
  \bibinfo {author} {\bibfnamefont {T.}~\bibnamefont {Tu}}, \bibinfo {author}
  {\bibfnamefont {L.}~\bibnamefont {Wang}}, \bibinfo {author} {\bibfnamefont
  {C.}~\bibnamefont {Zhou}}, \bibinfo {author} {\bibfnamefont {M.}~\bibnamefont
  {Xiao}}, \bibinfo {author} {\bibfnamefont {G.-C.}\ \bibnamefont {Guo}},
  \bibinfo {author} {\bibfnamefont {H.-W.}\ \bibnamefont {Jiang}},\ and\
  \bibinfo {author} {\bibfnamefont {G.-P.}\ \bibnamefont {Guo}},\ }\bibfield
  {title} {\bibinfo {title} {Ultrafast universal quantum control of a
  quantum-dot charge qubit using {L}andau--{Z}ener--{S}t{\"u}ckelberg
  interference},\ }\href {https://doi.org/10.1038/ncomms2412} {\bibfield
  {journal} {\bibinfo  {journal} {Nat. Commun.}\ }\textbf {\bibinfo {volume}
  {4}},\ \bibinfo {pages} {1401} (\bibinfo {year} {2013})}\BibitemShut
  {NoStop}%
\bibitem [{\citenamefont {Petersson}\ \emph {et~al.}(2010)\citenamefont
  {Petersson}, \citenamefont {Petta}, \citenamefont {Lu},\ and\ \citenamefont
  {Gossard}}]{petersson_2010}%
  \BibitemOpen
  \bibfield  {author} {\bibinfo {author} {\bibfnamefont {K.~D.}\ \bibnamefont
  {Petersson}}, \bibinfo {author} {\bibfnamefont {J.~R.}\ \bibnamefont
  {Petta}}, \bibinfo {author} {\bibfnamefont {H.}~\bibnamefont {Lu}},\ and\
  \bibinfo {author} {\bibfnamefont {A.~C.}\ \bibnamefont {Gossard}},\
  }\bibfield  {title} {\bibinfo {title} {Quantum coherence in a one-electron
  semiconductor charge qubit},\ }\href
  {https://doi.org/10.1103/PhysRevLett.105.246804} {\bibfield  {journal}
  {\bibinfo  {journal} {Phys. Rev. Lett.}\ }\textbf {\bibinfo {volume} {105}},\
  \bibinfo {pages} {246804} (\bibinfo {year} {2010})}\BibitemShut {NoStop}%
\bibitem [{\citenamefont {Li}\ \emph {et~al.}(2015)\citenamefont {Li},
  \citenamefont {Cao}, \citenamefont {Yu}, \citenamefont {Xiao}, \citenamefont
  {Guo}, \citenamefont {Jiang},\ and\ \citenamefont {Guo}}]{li_2015}%
  \BibitemOpen
  \bibfield  {author} {\bibinfo {author} {\bibfnamefont {H.-O.}\ \bibnamefont
  {Li}}, \bibinfo {author} {\bibfnamefont {G.}~\bibnamefont {Cao}}, \bibinfo
  {author} {\bibfnamefont {G.-D.}\ \bibnamefont {Yu}}, \bibinfo {author}
  {\bibfnamefont {M.}~\bibnamefont {Xiao}}, \bibinfo {author} {\bibfnamefont
  {G.-C.}\ \bibnamefont {Guo}}, \bibinfo {author} {\bibfnamefont {H.-W.}\
  \bibnamefont {Jiang}},\ and\ \bibinfo {author} {\bibfnamefont {G.-P.}\
  \bibnamefont {Guo}},\ }\bibfield  {title} {\bibinfo {title} {Conditional
  rotation of two strongly coupled semiconductor charge qubits},\ }\href
  {https://doi.org/10.1038/ncomms8681} {\bibfield  {journal} {\bibinfo
  {journal} {Nat. Commun.}\ }\textbf {\bibinfo {volume} {6}},\ \bibinfo {pages}
  {7681} (\bibinfo {year} {2015})}\BibitemShut {NoStop}%
\bibitem [{\citenamefont {Shinkai}\ \emph {et~al.}(2009)\citenamefont
  {Shinkai}, \citenamefont {Hayashi}, \citenamefont {Ota},\ and\ \citenamefont
  {Fujisawa}}]{shinkai_2009}%
  \BibitemOpen
  \bibfield  {author} {\bibinfo {author} {\bibfnamefont {G.}~\bibnamefont
  {Shinkai}}, \bibinfo {author} {\bibfnamefont {T.}~\bibnamefont {Hayashi}},
  \bibinfo {author} {\bibfnamefont {T.}~\bibnamefont {Ota}},\ and\ \bibinfo
  {author} {\bibfnamefont {T.}~\bibnamefont {Fujisawa}},\ }\bibfield  {title}
  {\bibinfo {title} {Correlated coherent oscillations in coupled semiconductor
  charge qubits},\ }\href {https://doi.org/10.1103/PhysRevLett.103.056802}
  {\bibfield  {journal} {\bibinfo  {journal} {Phys. Rev. Lett.}\ }\textbf
  {\bibinfo {volume} {103}},\ \bibinfo {pages} {056802} (\bibinfo {year}
  {2009})}\BibitemShut {NoStop}%
\bibitem [{\citenamefont {Vandersypen}\ \emph {et~al.}(2004)\citenamefont
  {Vandersypen}, \citenamefont {Elzerman}, \citenamefont {Schouten},
  \citenamefont {Willems~van Beveren}, \citenamefont {Hanson},\ and\
  \citenamefont {Kouwenhoven}}]{vandersypen_2004}%
  \BibitemOpen
  \bibfield  {author} {\bibinfo {author} {\bibfnamefont {L.~M.~K.}\
  \bibnamefont {Vandersypen}}, \bibinfo {author} {\bibfnamefont {J.~M.}\
  \bibnamefont {Elzerman}}, \bibinfo {author} {\bibfnamefont {R.~N.}\
  \bibnamefont {Schouten}}, \bibinfo {author} {\bibfnamefont {L.~H.}\
  \bibnamefont {Willems~van Beveren}}, \bibinfo {author} {\bibfnamefont
  {R.}~\bibnamefont {Hanson}},\ and\ \bibinfo {author} {\bibfnamefont {L.~P.}\
  \bibnamefont {Kouwenhoven}},\ }\bibfield  {title} {\bibinfo {title}
  {Real-time detection of single-electron tunneling using a quantum point
  contact},\ }\href {https://doi.org/10.1063/1.1815041} {\bibfield  {journal}
  {\bibinfo  {journal} {Appl. Phys. Lett.}\ }\textbf {\bibinfo {volume} {85}},\
  \bibinfo {pages} {4394} (\bibinfo {year} {2004})}\BibitemShut {NoStop}%
\bibitem [{\citenamefont {Gustavsson}\ \emph {et~al.}(2009)\citenamefont
  {Gustavsson}, \citenamefont {Leturcq}, \citenamefont {Studer}, \citenamefont
  {Shorubalko}, \citenamefont {Ihn}, \citenamefont {Ensslin}, \citenamefont
  {Driscoll},\ and\ \citenamefont {Gossard}}]{gustavsson_2009}%
  \BibitemOpen
  \bibfield  {author} {\bibinfo {author} {\bibfnamefont {S.}~\bibnamefont
  {Gustavsson}}, \bibinfo {author} {\bibfnamefont {R.}~\bibnamefont {Leturcq}},
  \bibinfo {author} {\bibfnamefont {M.}~\bibnamefont {Studer}}, \bibinfo
  {author} {\bibfnamefont {I.}~\bibnamefont {Shorubalko}}, \bibinfo {author}
  {\bibfnamefont {T.}~\bibnamefont {Ihn}}, \bibinfo {author} {\bibfnamefont
  {K.}~\bibnamefont {Ensslin}}, \bibinfo {author} {\bibfnamefont
  {D.}~\bibnamefont {Driscoll}},\ and\ \bibinfo {author} {\bibfnamefont
  {A.}~\bibnamefont {Gossard}},\ }\bibfield  {title} {\bibinfo {title}
  {Electron counting in quantum dots},\ }\href
  {https://doi.org/https://doi.org/10.1016/j.surfrep.2009.02.001} {\bibfield
  {journal} {\bibinfo  {journal} {Surf. Sci. Rep.}\ }\textbf {\bibinfo {volume}
  {64}},\ \bibinfo {pages} {191} (\bibinfo {year} {2009})}\BibitemShut
  {NoStop}%
\bibitem [{\citenamefont {Fujisawa}\ \emph {et~al.}(2004)\citenamefont
  {Fujisawa}, \citenamefont {Hayashi}, \citenamefont {Hirayama}, \citenamefont
  {Cheong},\ and\ \citenamefont {Jeong}}]{fujisawa_2004}%
  \BibitemOpen
  \bibfield  {author} {\bibinfo {author} {\bibfnamefont {T.}~\bibnamefont
  {Fujisawa}}, \bibinfo {author} {\bibfnamefont {T.}~\bibnamefont {Hayashi}},
  \bibinfo {author} {\bibfnamefont {Y.}~\bibnamefont {Hirayama}}, \bibinfo
  {author} {\bibfnamefont {H.~D.}\ \bibnamefont {Cheong}},\ and\ \bibinfo
  {author} {\bibfnamefont {Y.~H.}\ \bibnamefont {Jeong}},\ }\bibfield  {title}
  {\bibinfo {title} {Electron counting of single-electron tunneling current},\
  }\href {https://doi.org/10.1063/1.1691491} {\bibfield  {journal} {\bibinfo
  {journal} {Appl. Phys. Lett.}\ }\textbf {\bibinfo {volume} {84}},\ \bibinfo
  {pages} {2343} (\bibinfo {year} {2004})}\BibitemShut {NoStop}%
\bibitem [{\citenamefont {Lu}\ \emph {et~al.}(2003)\citenamefont {Lu},
  \citenamefont {Ji}, \citenamefont {Pfeiffer}, \citenamefont {West},\ and\
  \citenamefont {Rimberg}}]{lu_2003}%
  \BibitemOpen
  \bibfield  {author} {\bibinfo {author} {\bibfnamefont {W.}~\bibnamefont
  {Lu}}, \bibinfo {author} {\bibfnamefont {Z.}~\bibnamefont {Ji}}, \bibinfo
  {author} {\bibfnamefont {L.}~\bibnamefont {Pfeiffer}}, \bibinfo {author}
  {\bibfnamefont {K.~W.}\ \bibnamefont {West}},\ and\ \bibinfo {author}
  {\bibfnamefont {A.~J.}\ \bibnamefont {Rimberg}},\ }\bibfield  {title}
  {\bibinfo {title} {Real-time detection of electron tunnelling in a quantum
  dot},\ }\href {https://doi.org/10.1038/nature01642} {\bibfield  {journal}
  {\bibinfo  {journal} {Nature}\ }\textbf {\bibinfo {volume} {423}},\ \bibinfo
  {pages} {422} (\bibinfo {year} {2003})}\BibitemShut {NoStop}%
\bibitem [{\citenamefont {Rajabi}\ \emph {et~al.}(2013)\citenamefont {Rajabi},
  \citenamefont {P\"oltl},\ and\ \citenamefont {Governale}}]{rajabi_2013}%
  \BibitemOpen
  \bibfield  {author} {\bibinfo {author} {\bibfnamefont {L.}~\bibnamefont
  {Rajabi}}, \bibinfo {author} {\bibfnamefont {C.}~\bibnamefont {P\"oltl}},\
  and\ \bibinfo {author} {\bibfnamefont {M.}~\bibnamefont {Governale}},\
  }\bibfield  {title} {\bibinfo {title} {Waiting time distributions for the
  transport through a quantum-dot tunnel coupled to one normal and one
  superconducting lead},\ }\href
  {https://doi.org/10.1103/PhysRevLett.111.067002} {\bibfield  {journal}
  {\bibinfo  {journal} {Phys. Rev. Lett.}\ }\textbf {\bibinfo {volume} {111}},\
  \bibinfo {pages} {067002} (\bibinfo {year} {2013})}\BibitemShut {NoStop}%
\bibitem [{\citenamefont {Haack}\ \emph {et~al.}(2014)\citenamefont {Haack},
  \citenamefont {Albert},\ and\ \citenamefont {Flindt}}]{haack_2014}%
  \BibitemOpen
  \bibfield  {author} {\bibinfo {author} {\bibfnamefont {G.}~\bibnamefont
  {Haack}}, \bibinfo {author} {\bibfnamefont {M.}~\bibnamefont {Albert}},\ and\
  \bibinfo {author} {\bibfnamefont {C.}~\bibnamefont {Flindt}},\ }\bibfield
  {title} {\bibinfo {title} {Distributions of electron waiting times in
  quantum-coherent conductors},\ }\href
  {https://doi.org/10.1103/PhysRevB.90.205429} {\bibfield  {journal} {\bibinfo
  {journal} {Phys. Rev. B}\ }\textbf {\bibinfo {volume} {90}},\ \bibinfo
  {pages} {205429} (\bibinfo {year} {2014})}\BibitemShut {NoStop}%
\bibitem [{\citenamefont {Sothmann}(2014)}]{sothmann_2014}%
  \BibitemOpen
  \bibfield  {author} {\bibinfo {author} {\bibfnamefont {B.}~\bibnamefont
  {Sothmann}},\ }\bibfield  {title} {\bibinfo {title} {Electronic waiting-time
  distribution of a quantum-dot spin valve},\ }\href
  {https://doi.org/10.1103/PhysRevB.90.155315} {\bibfield  {journal} {\bibinfo
  {journal} {Phys. Rev. B}\ }\textbf {\bibinfo {volume} {90}},\ \bibinfo
  {pages} {155315} (\bibinfo {year} {2014})}\BibitemShut {NoStop}%
\bibitem [{\citenamefont {Kosov}(2017{\natexlab{a}})}]{kosov_2017}%
  \BibitemOpen
  \bibfield  {author} {\bibinfo {author} {\bibfnamefont {D.~S.}\ \bibnamefont
  {Kosov}},\ }\bibfield  {title} {\bibinfo {title} {Waiting time distribution
  for electron transport in a molecular junction with electron-vibration
  interaction},\ }\href {https://doi.org/10.1063/1.4976561} {\bibfield
  {journal} {\bibinfo  {journal} {J. Chem. Phys.}\ }\textbf {\bibinfo {volume}
  {146}},\ \bibinfo {pages} {074102} (\bibinfo {year}
  {2017}{\natexlab{a}})}\BibitemShut {NoStop}%
\bibitem [{\citenamefont {Kosov}(2017{\natexlab{b}})}]{kosov_2017_2}%
  \BibitemOpen
  \bibfield  {author} {\bibinfo {author} {\bibfnamefont {D.~S.}\ \bibnamefont
  {Kosov}},\ }\bibfield  {title} {\bibinfo {title} {Non-renewal statistics for
  electron transport in a molecular junction with electron-vibration
  interaction},\ }\href {https://doi.org/10.1063/1.4991038} {\bibfield
  {journal} {\bibinfo  {journal} {J. Chem. Phys.}\ }\textbf {\bibinfo {volume}
  {147}},\ \bibinfo {pages} {104109} (\bibinfo {year}
  {2017}{\natexlab{b}})}\BibitemShut {NoStop}%
\bibitem [{\citenamefont {Potanina}\ and\ \citenamefont
  {Flindt}(2017)}]{potanina_2017}%
  \BibitemOpen
  \bibfield  {author} {\bibinfo {author} {\bibfnamefont {E.}~\bibnamefont
  {Potanina}}\ and\ \bibinfo {author} {\bibfnamefont {C.}~\bibnamefont
  {Flindt}},\ }\bibfield  {title} {\bibinfo {title} {Electron waiting times of
  a periodically driven single-electron turnstile},\ }\href
  {https://doi.org/10.1103/PhysRevB.96.045420} {\bibfield  {journal} {\bibinfo
  {journal} {Phys. Rev. B}\ }\textbf {\bibinfo {volume} {96}},\ \bibinfo
  {pages} {045420} (\bibinfo {year} {2017})}\BibitemShut {NoStop}%
\bibitem [{\citenamefont {Rudge}\ and\ \citenamefont
  {Kosov}(2018)}]{rudge_2018}%
  \BibitemOpen
  \bibfield  {author} {\bibinfo {author} {\bibfnamefont {S.~L.}\ \bibnamefont
  {Rudge}}\ and\ \bibinfo {author} {\bibfnamefont {D.~S.}\ \bibnamefont
  {Kosov}},\ }\bibfield  {title} {\bibinfo {title} {Distribution of waiting
  times between electron cotunneling events},\ }\href
  {https://doi.org/10.1103/PhysRevB.98.245402} {\bibfield  {journal} {\bibinfo
  {journal} {Phys. Rev. B}\ }\textbf {\bibinfo {volume} {98}},\ \bibinfo
  {pages} {245402} (\bibinfo {year} {2018})}\BibitemShut {NoStop}%
\bibitem [{\citenamefont {Walldorf}\ \emph {et~al.}(2018)\citenamefont
  {Walldorf}, \citenamefont {Padurariu}, \citenamefont {Jauho},\ and\
  \citenamefont {Flindt}}]{walldorf_2018}%
  \BibitemOpen
  \bibfield  {author} {\bibinfo {author} {\bibfnamefont {N.}~\bibnamefont
  {Walldorf}}, \bibinfo {author} {\bibfnamefont {C.}~\bibnamefont {Padurariu}},
  \bibinfo {author} {\bibfnamefont {A.-P.}\ \bibnamefont {Jauho}},\ and\
  \bibinfo {author} {\bibfnamefont {C.}~\bibnamefont {Flindt}},\ }\bibfield
  {title} {\bibinfo {title} {Electron waiting times of a {C}ooper pair
  splitter},\ }\href {https://doi.org/10.1103/PhysRevLett.120.087701}
  {\bibfield  {journal} {\bibinfo  {journal} {Phys. Rev. Lett.}\ }\textbf
  {\bibinfo {volume} {120}},\ \bibinfo {pages} {087701} (\bibinfo {year}
  {2018})}\BibitemShut {NoStop}%
\bibitem [{\citenamefont {Tang}\ \emph {et~al.}(2018)\citenamefont {Tang},
  \citenamefont {Xu}, \citenamefont {Mi},\ and\ \citenamefont
  {Wang}}]{tang_2018}%
  \BibitemOpen
  \bibfield  {author} {\bibinfo {author} {\bibfnamefont {G.}~\bibnamefont
  {Tang}}, \bibinfo {author} {\bibfnamefont {F.}~\bibnamefont {Xu}}, \bibinfo
  {author} {\bibfnamefont {S.}~\bibnamefont {Mi}},\ and\ \bibinfo {author}
  {\bibfnamefont {J.}~\bibnamefont {Wang}},\ }\bibfield  {title} {\bibinfo
  {title} {Spin-resolved electron waiting times in a quantum-dot spin valve},\
  }\href {https://doi.org/10.1103/PhysRevB.97.165407} {\bibfield  {journal}
  {\bibinfo  {journal} {Phys. Rev. B}\ }\textbf {\bibinfo {volume} {97}},\
  \bibinfo {pages} {165407} (\bibinfo {year} {2018})}\BibitemShut {NoStop}%
\bibitem [{\citenamefont {Engelhardt}\ and\ \citenamefont
  {Cao}(2019)}]{engelhardt_2019}%
  \BibitemOpen
  \bibfield  {author} {\bibinfo {author} {\bibfnamefont {G.}~\bibnamefont
  {Engelhardt}}\ and\ \bibinfo {author} {\bibfnamefont {J.}~\bibnamefont
  {Cao}},\ }\bibfield  {title} {\bibinfo {title} {Tuning the {A}haronov-{B}ohm
  effect with dephasing in nonequilibrium transport},\ }\href
  {https://doi.org/10.1103/PhysRevB.99.075436} {\bibfield  {journal} {\bibinfo
  {journal} {Phys. Rev. B}\ }\textbf {\bibinfo {volume} {99}},\ \bibinfo
  {pages} {075436} (\bibinfo {year} {2019})}\BibitemShut {NoStop}%
\bibitem [{\citenamefont {Rudge}\ and\ \citenamefont
  {Kosov}(2019)}]{rudge_2019}%
  \BibitemOpen
  \bibfield  {author} {\bibinfo {author} {\bibfnamefont {S.~L.}\ \bibnamefont
  {Rudge}}\ and\ \bibinfo {author} {\bibfnamefont {D.~S.}\ \bibnamefont
  {Kosov}},\ }\bibfield  {title} {\bibinfo {title} {Counting quantum jumps: A
  summary and comparison of fixed-time and fluctuating-time statistics in
  electron transport},\ }\href {https://doi.org/10.1063/1.5108518} {\bibfield
  {journal} {\bibinfo  {journal} {J. Chem. Phys.}\ }\textbf {\bibinfo {volume}
  {151}},\ \bibinfo {pages} {034107} (\bibinfo {year} {2019})}\BibitemShut
  {NoStop}%
\bibitem [{\citenamefont {Stegmann}\ \emph {et~al.}(2021)\citenamefont
  {Stegmann}, \citenamefont {Sothmann}, \citenamefont {K\"onig},\ and\
  \citenamefont {Flindt}}]{stegmann_2021}%
  \BibitemOpen
  \bibfield  {author} {\bibinfo {author} {\bibfnamefont {P.}~\bibnamefont
  {Stegmann}}, \bibinfo {author} {\bibfnamefont {B.}~\bibnamefont {Sothmann}},
  \bibinfo {author} {\bibfnamefont {J.}~\bibnamefont {K\"onig}},\ and\ \bibinfo
  {author} {\bibfnamefont {C.}~\bibnamefont {Flindt}},\ }\bibfield  {title}
  {\bibinfo {title} {Electron waiting times in a strongly interacting quantum
  dot: Interaction effects and higher-order tunneling processes},\ }\href
  {https://doi.org/10.1103/PhysRevLett.127.096803} {\bibfield  {journal}
  {\bibinfo  {journal} {Phys. Rev. Lett.}\ }\textbf {\bibinfo {volume} {127}},\
  \bibinfo {pages} {096803} (\bibinfo {year} {2021})}\BibitemShut {NoStop}%
\bibitem [{\citenamefont {Davis}\ \emph {et~al.}(2021)\citenamefont {Davis},
  \citenamefont {Rudge},\ and\ \citenamefont {Kosov}}]{davis_2021}%
  \BibitemOpen
  \bibfield  {author} {\bibinfo {author} {\bibfnamefont {N.~S.}\ \bibnamefont
  {Davis}}, \bibinfo {author} {\bibfnamefont {S.~L.}\ \bibnamefont {Rudge}},\
  and\ \bibinfo {author} {\bibfnamefont {D.~S.}\ \bibnamefont {Kosov}},\
  }\bibfield  {title} {\bibinfo {title} {Electronic statistics on demand:
  Bunching, antibunching, positive, and negative correlations in a molecular
  spin valve},\ }\href {https://doi.org/10.1103/PhysRevB.103.205408} {\bibfield
   {journal} {\bibinfo  {journal} {Phys. Rev. B}\ }\textbf {\bibinfo {volume}
  {103}},\ \bibinfo {pages} {205408} (\bibinfo {year} {2021})}\BibitemShut
  {NoStop}%
\bibitem [{\citenamefont {Landi}()}]{landi_2021}%
  \BibitemOpen
  \bibfield  {author} {\bibinfo {author} {\bibfnamefont {G.~T.}\ \bibnamefont
  {Landi}},\ }\href@noop {} {\bibinfo {title} {Waiting-times statistics in
  boundary driven free fermion chains}},\ \Eprint
  {https://arxiv.org/abs/2108.11850} {arXiv:2108.11850 [quant-ph]} \BibitemShut
  {NoStop}%
\bibitem [{\citenamefont {Emary}\ \emph {et~al.}(2012)\citenamefont {Emary},
  \citenamefont {P\"oltl}, \citenamefont {Carmele}, \citenamefont {Kabuss},
  \citenamefont {Knorr},\ and\ \citenamefont {Brandes}}]{emary_2012}%
  \BibitemOpen
  \bibfield  {author} {\bibinfo {author} {\bibfnamefont {C.}~\bibnamefont
  {Emary}}, \bibinfo {author} {\bibfnamefont {C.}~\bibnamefont {P\"oltl}},
  \bibinfo {author} {\bibfnamefont {A.}~\bibnamefont {Carmele}}, \bibinfo
  {author} {\bibfnamefont {J.}~\bibnamefont {Kabuss}}, \bibinfo {author}
  {\bibfnamefont {A.}~\bibnamefont {Knorr}},\ and\ \bibinfo {author}
  {\bibfnamefont {T.}~\bibnamefont {Brandes}},\ }\bibfield  {title} {\bibinfo
  {title} {Bunching and antibunching in electronic transport},\ }\href
  {https://doi.org/10.1103/PhysRevB.85.165417} {\bibfield  {journal} {\bibinfo
  {journal} {Phys. Rev. B}\ }\textbf {\bibinfo {volume} {85}},\ \bibinfo
  {pages} {165417} (\bibinfo {year} {2012})}\BibitemShut {NoStop}%
\bibitem [{\citenamefont {Ptaszy\ifmmode~\acute{n}\else
  \'{n}\fi{}ski}(2017)}]{ptaszynski_2017}%
  \BibitemOpen
  \bibfield  {author} {\bibinfo {author} {\bibfnamefont {K.}~\bibnamefont
  {Ptaszy\ifmmode~\acute{n}\else \'{n}\fi{}ski}},\ }\bibfield  {title}
  {\bibinfo {title} {Waiting time distribution revealing the internal spin
  dynamics in a double quantum dot},\ }\href
  {https://doi.org/10.1103/PhysRevB.96.035409} {\bibfield  {journal} {\bibinfo
  {journal} {Phys. Rev. B}\ }\textbf {\bibinfo {volume} {96}},\ \bibinfo
  {pages} {035409} (\bibinfo {year} {2017})}\BibitemShut {NoStop}%
\bibitem [{\citenamefont {Thomas}\ and\ \citenamefont
  {Flindt}(2014)}]{thomas_2014}%
  \BibitemOpen
  \bibfield  {author} {\bibinfo {author} {\bibfnamefont {K.~H.}\ \bibnamefont
  {Thomas}}\ and\ \bibinfo {author} {\bibfnamefont {C.}~\bibnamefont
  {Flindt}},\ }\bibfield  {title} {\bibinfo {title} {Waiting time distributions
  of noninteracting fermions on a tight-binding chain},\ }\href
  {https://doi.org/10.1103/PhysRevB.89.245420} {\bibfield  {journal} {\bibinfo
  {journal} {Phys. Rev. B}\ }\textbf {\bibinfo {volume} {89}},\ \bibinfo
  {pages} {245420} (\bibinfo {year} {2014})}\BibitemShut {NoStop}%
\bibitem [{\citenamefont {Kleinherbers}\ \emph {et~al.}(2020)\citenamefont
  {Kleinherbers}, \citenamefont {Szpak}, \citenamefont {K\"onig},\ and\
  \citenamefont {Sch\"utzhold}}]{kleinherbers_2020}%
  \BibitemOpen
  \bibfield  {author} {\bibinfo {author} {\bibfnamefont {E.}~\bibnamefont
  {Kleinherbers}}, \bibinfo {author} {\bibfnamefont {N.}~\bibnamefont {Szpak}},
  \bibinfo {author} {\bibfnamefont {J.}~\bibnamefont {K\"onig}},\ and\ \bibinfo
  {author} {\bibfnamefont {R.}~\bibnamefont {Sch\"utzhold}},\ }\bibfield
  {title} {\bibinfo {title} {Relaxation dynamics in a {H}ubbard dimer coupled
  to fermionic baths: Phenomenological description and its microscopic
  foundation},\ }\href {https://doi.org/10.1103/PhysRevB.101.125131} {\bibfield
   {journal} {\bibinfo  {journal} {Phys. Rev. B}\ }\textbf {\bibinfo {volume}
  {101}},\ \bibinfo {pages} {125131} (\bibinfo {year} {2020})}\BibitemShut
  {NoStop}%
\bibitem [{\citenamefont {Kir\ifmmode~\check{s}\else \v{s}\fi{}anskas}\ \emph
  {et~al.}(2018)\citenamefont {Kir\ifmmode~\check{s}\else \v{s}\fi{}anskas},
  \citenamefont {Francki\'e},\ and\ \citenamefont {Wacker}}]{gediminas_2018}%
  \BibitemOpen
  \bibfield  {author} {\bibinfo {author} {\bibfnamefont {G.}~\bibnamefont
  {Kir\ifmmode~\check{s}\else \v{s}\fi{}anskas}}, \bibinfo {author}
  {\bibfnamefont {M.}~\bibnamefont {Francki\'e}},\ and\ \bibinfo {author}
  {\bibfnamefont {A.}~\bibnamefont {Wacker}},\ }\bibfield  {title} {\bibinfo
  {title} {Phenomenological position and energy resolving {L}indblad approach
  to quantum kinetics},\ }\href {https://doi.org/10.1103/PhysRevB.97.035432}
  {\bibfield  {journal} {\bibinfo  {journal} {Phys. Rev. B}\ }\textbf {\bibinfo
  {volume} {97}},\ \bibinfo {pages} {035432} (\bibinfo {year}
  {2018})}\BibitemShut {NoStop}%
\bibitem [{\citenamefont {Ptaszy\ifmmode~\acute{n}\else \'{n}\fi{}ski}\ and\
  \citenamefont {Esposito}(2019)}]{ptaszynski_2019}%
  \BibitemOpen
  \bibfield  {author} {\bibinfo {author} {\bibfnamefont {K.}~\bibnamefont
  {Ptaszy\ifmmode~\acute{n}\else \'{n}\fi{}ski}}\ and\ \bibinfo {author}
  {\bibfnamefont {M.}~\bibnamefont {Esposito}},\ }\bibfield  {title} {\bibinfo
  {title} {Thermodynamics of quantum information flows},\ }\href
  {https://doi.org/10.1103/PhysRevLett.122.150603} {\bibfield  {journal}
  {\bibinfo  {journal} {Phys. Rev. Lett.}\ }\textbf {\bibinfo {volume} {122}},\
  \bibinfo {pages} {150603} (\bibinfo {year} {2019})}\BibitemShut {NoStop}%
\bibitem [{\citenamefont {Timm}(2008)}]{timm_2008}%
  \BibitemOpen
  \bibfield  {author} {\bibinfo {author} {\bibfnamefont {C.}~\bibnamefont
  {Timm}},\ }\bibfield  {title} {\bibinfo {title} {Tunneling through molecules
  and quantum dots: Master-equation approaches},\ }\href
  {https://doi.org/10.1103/PhysRevB.77.195416} {\bibfield  {journal} {\bibinfo
  {journal} {Phys. Rev. B}\ }\textbf {\bibinfo {volume} {77}},\ \bibinfo
  {pages} {195416} (\bibinfo {year} {2008})}\BibitemShut {NoStop}%
\bibitem [{\citenamefont {Wunsch}\ \emph {et~al.}(2005)\citenamefont {Wunsch},
  \citenamefont {Braun}, \citenamefont {K\"onig},\ and\ \citenamefont
  {Pfannkuche}}]{wunsch_2005}%
  \BibitemOpen
  \bibfield  {author} {\bibinfo {author} {\bibfnamefont {B.}~\bibnamefont
  {Wunsch}}, \bibinfo {author} {\bibfnamefont {M.}~\bibnamefont {Braun}},
  \bibinfo {author} {\bibfnamefont {J.}~\bibnamefont {K\"onig}},\ and\ \bibinfo
  {author} {\bibfnamefont {D.}~\bibnamefont {Pfannkuche}},\ }\bibfield  {title}
  {\bibinfo {title} {Probing level renormalization by sequential transport
  through double quantum dots},\ }\href
  {https://doi.org/10.1103/PhysRevB.72.205319} {\bibfield  {journal} {\bibinfo
  {journal} {Phys. Rev. B}\ }\textbf {\bibinfo {volume} {72}},\ \bibinfo
  {pages} {205319} (\bibinfo {year} {2005})}\BibitemShut {NoStop}%
\bibitem [{\citenamefont {Splettstoesser}\ \emph {et~al.}(2012)\citenamefont
  {Splettstoesser}, \citenamefont {Governale},\ and\ \citenamefont
  {K\"onig}}]{splettstoesser_2012}%
  \BibitemOpen
  \bibfield  {author} {\bibinfo {author} {\bibfnamefont {J.}~\bibnamefont
  {Splettstoesser}}, \bibinfo {author} {\bibfnamefont {M.}~\bibnamefont
  {Governale}},\ and\ \bibinfo {author} {\bibfnamefont {J.}~\bibnamefont
  {K\"onig}},\ }\bibfield  {title} {\bibinfo {title} {Tunneling-induced
  renormalization in interacting quantum dots},\ }\href
  {https://doi.org/10.1103/PhysRevB.86.035432} {\bibfield  {journal} {\bibinfo
  {journal} {Phys. Rev. B}\ }\textbf {\bibinfo {volume} {86}},\ \bibinfo
  {pages} {035432} (\bibinfo {year} {2012})}\BibitemShut {NoStop}%
\bibitem [{\citenamefont {Stegmann}\ \emph {et~al.}(2018)\citenamefont
  {Stegmann}, \citenamefont {K\"onig},\ and\ \citenamefont
  {Weiss}}]{stegmann_2018}%
  \BibitemOpen
  \bibfield  {author} {\bibinfo {author} {\bibfnamefont {P.}~\bibnamefont
  {Stegmann}}, \bibinfo {author} {\bibfnamefont {J.}~\bibnamefont {K\"onig}},\
  and\ \bibinfo {author} {\bibfnamefont {S.}~\bibnamefont {Weiss}},\ }\bibfield
   {title} {\bibinfo {title} {Coherent dynamics in stochastic systems revealed
  by full counting statistics},\ }\href
  {https://doi.org/10.1103/PhysRevB.98.035409} {\bibfield  {journal} {\bibinfo
  {journal} {Phys. Rev. B}\ }\textbf {\bibinfo {volume} {98}},\ \bibinfo
  {pages} {035409} (\bibinfo {year} {2018})}\BibitemShut {NoStop}%
\bibitem [{\citenamefont {Stegmann}\ \emph {et~al.}(2020)\citenamefont
  {Stegmann}, \citenamefont {K\"onig},\ and\ \citenamefont
  {Sothmann}}]{stegmann_2020_2}%
  \BibitemOpen
  \bibfield  {author} {\bibinfo {author} {\bibfnamefont {P.}~\bibnamefont
  {Stegmann}}, \bibinfo {author} {\bibfnamefont {J.}~\bibnamefont {K\"onig}},\
  and\ \bibinfo {author} {\bibfnamefont {B.}~\bibnamefont {Sothmann}},\
  }\bibfield  {title} {\bibinfo {title} {Relaxation dynamics in double-spin
  systems},\ }\href {https://doi.org/10.1103/PhysRevB.101.075411} {\bibfield
  {journal} {\bibinfo  {journal} {Phys. Rev. B}\ }\textbf {\bibinfo {volume}
  {101}},\ \bibinfo {pages} {075411} (\bibinfo {year} {2020})}\BibitemShut
  {NoStop}%
\bibitem [{\citenamefont {Ghoshal}\ and\ \citenamefont {Sen}()}]{ghoshal_2021}%
  \BibitemOpen
  \bibfield  {author} {\bibinfo {author} {\bibfnamefont {A.}~\bibnamefont
  {Ghoshal}}\ and\ \bibinfo {author} {\bibfnamefont {U.}~\bibnamefont {Sen}},\
  }\href@noop {} {\bibinfo {title} {Heat current and entropy production rate in
  local non-{M}arkovian quantum dynamics of global {M}arkovian evolution}},\
  \Eprint {https://arxiv.org/abs/2102.06694} {arXiv:2102.06694 [quant-ph]}
  \BibitemShut {NoStop}%
\bibitem [{\citenamefont {Roulet}\ and\ \citenamefont
  {Bruder}(2018{\natexlab{a}})}]{roulet_2018_1}%
  \BibitemOpen
  \bibfield  {author} {\bibinfo {author} {\bibfnamefont {A.}~\bibnamefont
  {Roulet}}\ and\ \bibinfo {author} {\bibfnamefont {C.}~\bibnamefont
  {Bruder}},\ }\bibfield  {title} {\bibinfo {title} {Quantum synchronization
  and entanglement generation},\ }\href
  {https://doi.org/10.1103/PhysRevLett.121.063601} {\bibfield  {journal}
  {\bibinfo  {journal} {Phys. Rev. Lett.}\ }\textbf {\bibinfo {volume} {121}},\
  \bibinfo {pages} {063601} (\bibinfo {year} {2018}{\natexlab{a}})}\BibitemShut
  {NoStop}%
\bibitem [{\citenamefont {Roulet}\ and\ \citenamefont
  {Bruder}(2018{\natexlab{b}})}]{roulet_2018_2}%
  \BibitemOpen
  \bibfield  {author} {\bibinfo {author} {\bibfnamefont {A.}~\bibnamefont
  {Roulet}}\ and\ \bibinfo {author} {\bibfnamefont {C.}~\bibnamefont
  {Bruder}},\ }\bibfield  {title} {\bibinfo {title} {Synchronizing the smallest
  possible system},\ }\href {https://doi.org/10.1103/PhysRevLett.121.053601}
  {\bibfield  {journal} {\bibinfo  {journal} {Phys. Rev. Lett.}\ }\textbf
  {\bibinfo {volume} {121}},\ \bibinfo {pages} {053601} (\bibinfo {year}
  {2018}{\natexlab{b}})}\BibitemShut {NoStop}%
\bibitem [{\citenamefont {Giorgi}\ \emph {et~al.}(2013)\citenamefont {Giorgi},
  \citenamefont {Plastina}, \citenamefont {Francica},\ and\ \citenamefont
  {Zambrini}}]{giorgi_2013}%
  \BibitemOpen
  \bibfield  {author} {\bibinfo {author} {\bibfnamefont {G.~L.}\ \bibnamefont
  {Giorgi}}, \bibinfo {author} {\bibfnamefont {F.}~\bibnamefont {Plastina}},
  \bibinfo {author} {\bibfnamefont {G.}~\bibnamefont {Francica}},\ and\
  \bibinfo {author} {\bibfnamefont {R.}~\bibnamefont {Zambrini}},\ }\bibfield
  {title} {\bibinfo {title} {Spontaneous synchronization and quantum
  correlation dynamics of open spin systems},\ }\href
  {https://doi.org/10.1103/PhysRevA.88.042115} {\bibfield  {journal} {\bibinfo
  {journal} {Phys. Rev. A}\ }\textbf {\bibinfo {volume} {88}},\ \bibinfo
  {pages} {042115} (\bibinfo {year} {2013})}\BibitemShut {NoStop}%
\bibitem [{\citenamefont {Dambach}\ \emph {et~al.}(2015)\citenamefont
  {Dambach}, \citenamefont {Kubala}, \citenamefont {Gramich},\ and\
  \citenamefont {Ankerhold}}]{dambach_2015}%
  \BibitemOpen
  \bibfield  {author} {\bibinfo {author} {\bibfnamefont {S.}~\bibnamefont
  {Dambach}}, \bibinfo {author} {\bibfnamefont {B.}~\bibnamefont {Kubala}},
  \bibinfo {author} {\bibfnamefont {V.}~\bibnamefont {Gramich}},\ and\ \bibinfo
  {author} {\bibfnamefont {J.}~\bibnamefont {Ankerhold}},\ }\bibfield  {title}
  {\bibinfo {title} {Time-resolved statistics of nonclassical light in
  {J}osephson photonics},\ }\href {https://doi.org/10.1103/PhysRevB.92.054508}
  {\bibfield  {journal} {\bibinfo  {journal} {Phys. Rev. B}\ }\textbf {\bibinfo
  {volume} {92}},\ \bibinfo {pages} {054508} (\bibinfo {year}
  {2015})}\BibitemShut {NoStop}%
\end{thebibliography}
\end{document}